\documentclass[aps,prb,twocolumn,superscriptaddress,english]{revtex4-1}

\usepackage[T1]{fontenc}
\usepackage{babel}
\usepackage{amsmath}
\usepackage{amssymb}
\usepackage{wasysym}
\usepackage{graphicx}
\usepackage{xcolor}
\usepackage{braket}
\usepackage{booktabs}
\usepackage{multirow}
\usepackage{makecell}
\usepackage{titlesec}
\usepackage{xcolor}

\usepackage[linktocpage=true,
  colorlinks=true, 
  pdfborder={0 0 0},
  linkcolor=blue,
  citecolor=red,
  filecolor=yellow,
  urlcolor=blue,
  bookmarks,
  pdfauthor={},
]{hyperref}

\newcommand{\lah}{LaH$_{10}$}

\newcommand{\tc}{$T_\text{c}$}

\newcommand{\omlog}{$\omega_{\textmd log}$}
\newcommand{\ep}{\textit{e-ph}~}

\titleformat{\paragraph}[runin]
{\bfseries}{\theparagraph}{1em}{}

\begin{document}

\title{La-$X$-H hydrides: is hot superconductivity possible?}

\author{Simone Di Cataldo} \email{simone.dicataldo@uniroma1.it}
\affiliation{Institute of Theoretical and Computational Physics, Graz University of Technology, NAWI Graz, 8010 Graz, Austria}
\affiliation{Dipartimento di Fisica, Sapienza Universit\`a di Roma, 00185 Roma, Italy} 
\author{Wolfgang von der Linden}
\affiliation{Institute of Theoretical and Computational Physics, Graz University of Technology, NAWI Graz, 8010 Graz, Austria}
\author{Lilia Boeri} \email{lilia.boeri@uniroma1.it}
\affiliation{Dipartimento di Fisica, Sapienza Universit\`a di Roma, 00185 Roma, Italy} 

\begin{abstract}
  Motivated by the recent claim of \textit{hot} superconductivity with
  critical temperatures up to 550 K in La + $x$ hydrides (arXiv:2006.03004),
  we investigate the high-pressure phase diagram of possible compounds that may have
  formed in the experiment,
  using first-principles calculations for evolutionary crystal structure
  prediction and superconductivity.
  Starting from the hypothesis that the observed \tc{} may be realized by successive heating
  upon a pre-formed LaH$_{10}$  phase,
  we examine plausible ternaries of lanthanum, hydrogen and other
  elements present in the diamond anvil cell: boron, nitrogen, carbon, platinum, , gallium, gold.
  We find that only boron forms superhydride-like structures that can host high-\tc{} superconductivity,
  but the predicted \tc{} are incompatibe with the experimental reports.
  Our results indicate that, while the claims of \textit{hot} superconductivity should be reconsidered,
  it is very likely that unkwown H-rich ternary or multinary phases
  containing lanthanum, boron and hydrogen may have formed under the reported experimental conditions,
  and that these may exhibit superconducting properties comparable, or even superior, to those of currently known hydrides.
\end{abstract}

\maketitle
%%\section{Introduction}
%\section{Main}
Since the discovery of high-temperature superconductivity (HTSC) in compressed
sulfur hydride in 2014 \cite{Eremets_Nature_2015_SH3, Duan_SciRep_2014_SH, Eremets_NatPhys_2016_SH3},
the race for high-temperature superconductivity has dramatically accelerated,
leading to a {\em hydride rush} fueled by \textit{ab-initio} predictions.

As of 2020, all binary hydrides have been computationally explored \cite{Zurek_JCP_2019_review, Boeri_PhysRep_2020_review, Oganov_SSM_2020_MH}, and many have been synthesized \cite{Eremets_Nature_2019_LaH, Hemley_PRL_2019_LaH, Ashcroft_PNAS_2017_LaH, Oganov_MatToday_2020_ThH, Oganov_arXiv_2019_YH6, Eremets_arXiv_2019_YH6, Eremets_arXiv_PH3_2015}.
After \lah{} established the \tc{} record for binary hydrides in 2019, in 2020 Snider et al. \cite{Dias_Nature_2020_CSH} 
reported a \tc{} of 288 K in a compressed mixture of carbon, sulfur and hydrogen,
effectively realizing the first room-temperature superconductor.
Compared to binary hydrides, ternary (or, in general, multinary) hydrides exhibit an
increased chemical flexibility, which may be exploited to tune the superconducting properties.
Since Migdal-Eliashberg theory does not pose a hard limit to $T_c$, it is 
possible that multinary hydrides may exhibit superconductivity at sensibly higher \tc{}s than the known binaries;  for example,
\tc's largely exceeding room temperature have been predicted in a Li-Mg-H alloy.\cite{Ma_PRL_2019_Li2MgH16}.

In the summer of 2020, Grockowiak et al. reported experimental evidence of superconductivity with onset temperatures growing from 294 K to 550 K upon successive heating cycles of a mixture of lanthanum and ammonia borane at about 180 GPa \cite{superconductivity_550K}.
This may have been the first experimental observation of \textit{hot superconductivity} in a multinary hydride;
unfortunately, due to COVID restrictions, the authors were able to report only partial
evidences, and did not provide information on the chemical composition and structure of the superconducting samples, which would be fundamental for reproducibility. Even if one is skeptical about the highest values of $T_c$ reported, it is possible that one or more new multinary phases may have formed, calling for further studies. 
In the absence of conclusive experimental information, such a question can effectively be addressed by first-principles methods for crystal structure prediction and superconductivity, which have demonstrated an extraordinary accuracy for binary hydrides.~\cite{Zurek_JCP_2019_review, Boeri_PhysRep_2020_review, Oganov_SSM_2020_MH, Boeri_JPCM_2019_viewpoint}

\begin{figure*}[ht]
	\centering
	\includegraphics[width=0.48\linewidth]{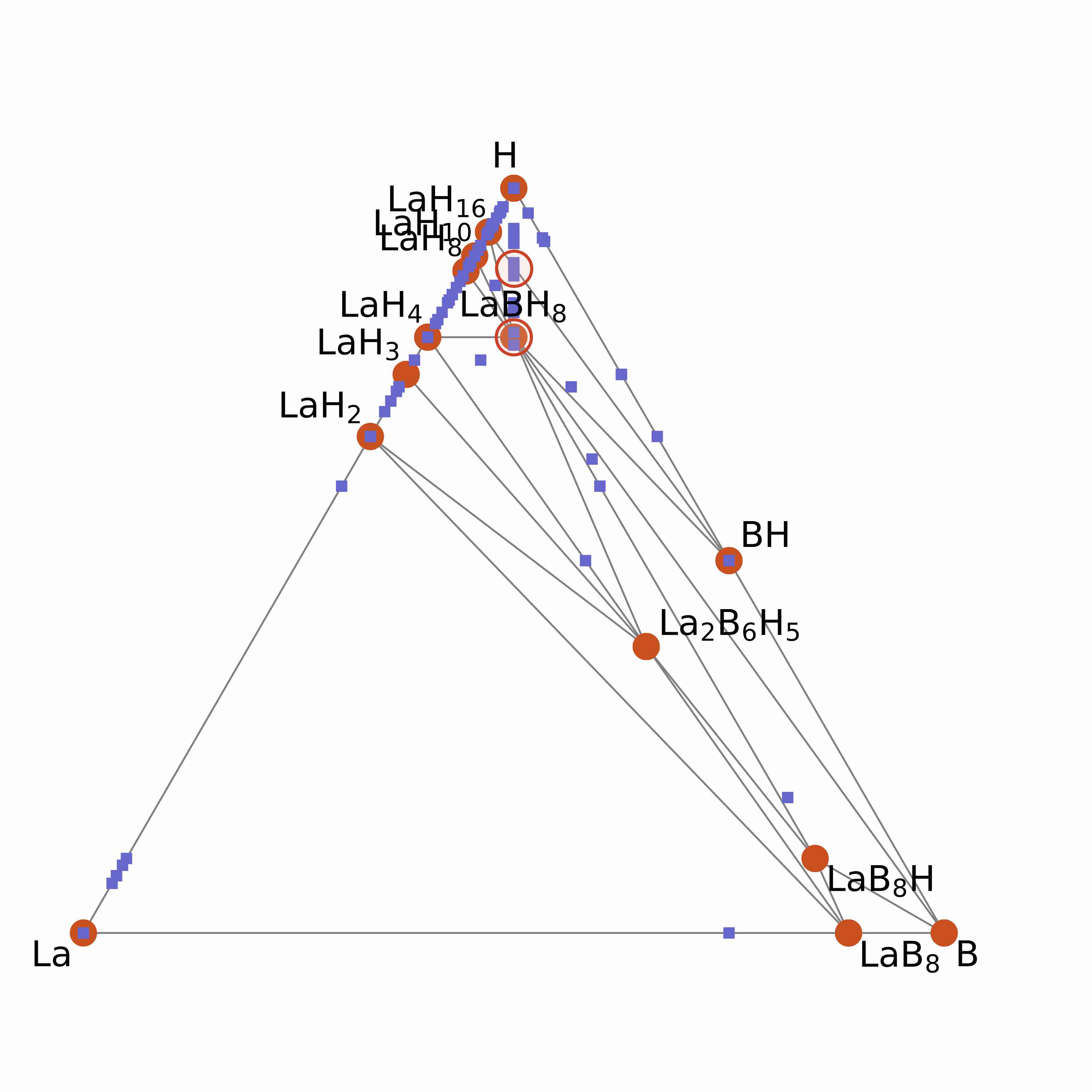}
	\includegraphics[width=0.48\linewidth]{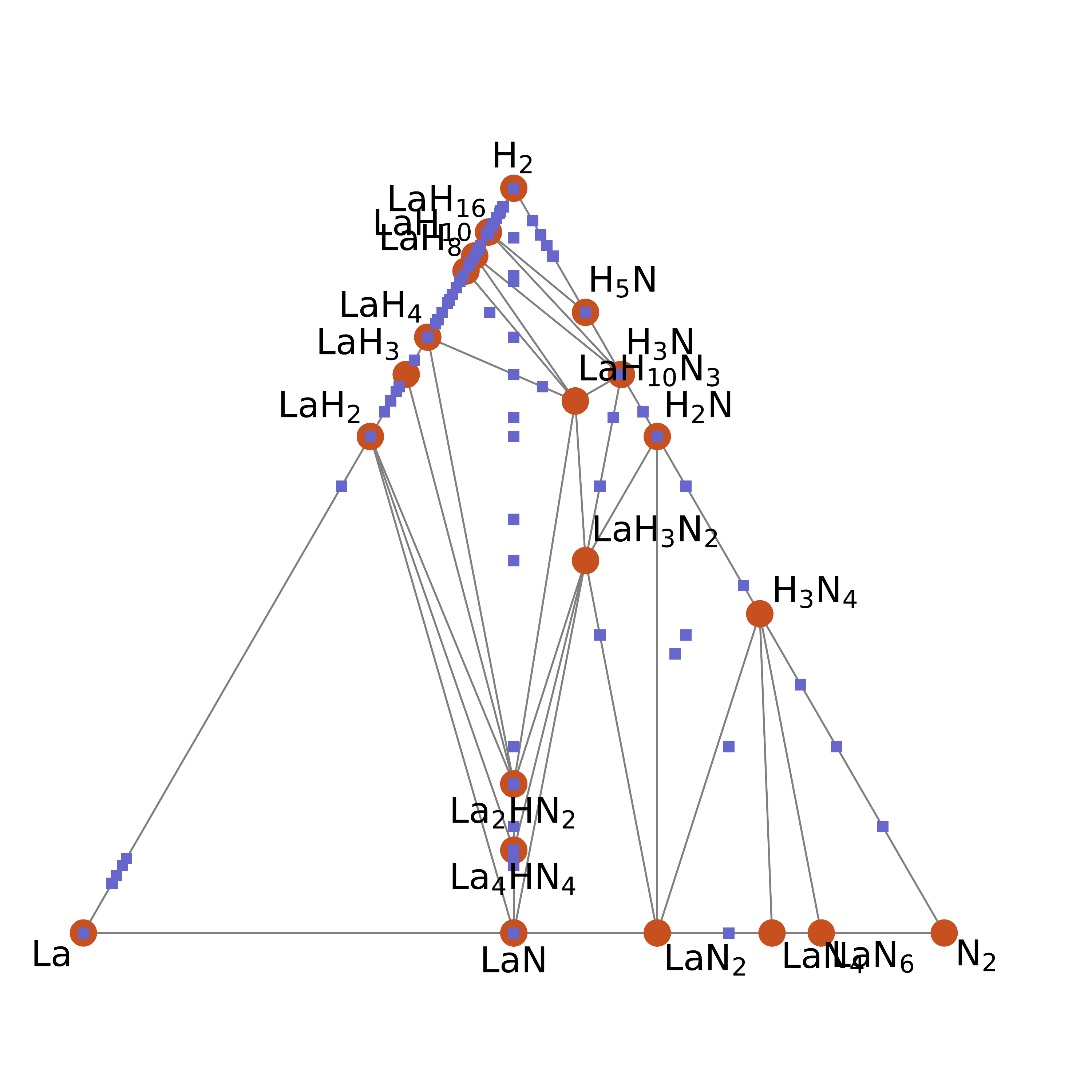}
	\caption{Convex hulls for La-N-H (left) and La-B-H (right) at 300 GPa. Orange circles and blue squares represent stable and metastable phases, respectively.}
	\label{fig:hulls}
\end{figure*}
In this paper, we perform an exploratory ab-initio study of possible candidates for hot superconductivity, using evolutionary crystal structure prediction and linear-response calculations of the electron-phonon coupling. We explore all possible ternary combinations of lanthanum, hydrogen, and
a third element present in the diamond anvil cell (DAC) in the experiment of Ref.~\onlinecite{superconductivity_550K}: boron and nitrogen (from the hydrogen source), carbon, from the epoxy, platinum, gallium and gold, from the electrical contacts. One (or more) of these elements may react with lanthanum and hydrogen to form a new, unknown superhydride.
Our aim is to sample the ternary phase diagrams
with an accuracy sufficient to estimate the probability that stable or weakly metastable
structures may form at high pressure, determine the characteristic structural motifs, 
and assess their potential for high-\tc{} conventional superconductivity. We choose to carry out all of our structural searches at 300 GPa, rather than 180 GPa, which is the highest pressure measured in the experiment.
In fact, there is rather extensive evidence that static Density Functional Theory (DFT) calculations, which neglect quantum lattice effects, tend to overestimate the stabilization pressure of hydrides\cite{Mauri_Nature_2016_SH3, Mauri_Nature_2020_LaH} by as much as 100 GPa, and higher pressures typically yield better agreement
with experiments also for superconducting properties.

We show that, among all elements present in the DAC, only boron forms ternary
structures with La and H with high \tc's; most elements do not form any ternary structure at high pressures (C, Pt, Ga, Au),
while nitrogen forms stable and metastable structures which do not exhibit the typical
characteristics of high-\tc{} superhydrides: high-symmetry, large hydrogen content, large fraction
of H states at the Fermi level.
On the other hand, some La-B-H structures are characterized by the same hydrogen cage-like motifs encountered in many record superhydrides \cite{Zurek_PNAS_2009_LiH, Ma_PRL_2017_ReH}.
In particular, within the limitations posed by the maximum cell size, our best superconducting phase is LaBH$_{17}$, with a \tc{} of 180 K, which is way too low to explain
the hot superconductivity observed in Ref.~\onlinecite{superconductivity_550K}.
Tuning of the electronic and vibrational properties of La-B-H structures through doping or impurities may 
increase this value up to a factor two, but it is extremely unlikely that this type of structures may reach \tc's as high as 550 K.

%%%
This paper is organized as follows: In sect. I, we discuss some general criteria leading to the formation of stable ternary phases, and present the predicted phase diagrams;
In sect. II, we analyze the electronic structure of the predicted phases. In sect. III, we discuss the superconducting properties of the metallic phases.

\section{Phase Diagrams}
Our structural searches were carried out using evolutionary algorithms as implemented in the Universal Structure Predictor: Evolutionary Xtallography code (USPEX)\cite{USPEX_1, USPEX_2}.
Further details on the structural searches can be found in the Supplemental Material.
Since full structural searches of ternary diagrams are extremely expensive computationally, before sampling
in-depth the ternary compositions, we carried out a pre-screening process.

%This section is organized as follows: first we are going to discuss the choice of elements to investigate, based on considerations on the experimental report; then we will discuss the results of our preliminary scans, whereby we determined that only La-B-H and La-N-H were worth further investigation; finally, we will report the full ternary phase diagram for La-B-H, and La-N-H. 

Our strategy is inspired by the work of Sun and coworkers \cite{Ceder_NatMat_2019}, where the authors thoroughly demonstrated that in ternary nitrides the \textit{depth} (i.e. the energy of the lowest-lying binary composition) of the associated binary convex hulls is a very good indicator of whether thermodynamically stable ternary phases will form.
A zero, or positive depth (no stable binary phases) indicates no chemical affinity, while a too large negative depth implies a strong thermochemical competition
of binary compositions, which destabilizes ternary phases.
Both conditions can prevent the formation of stable ternaries. The otpimal depth of the binary convex hulls is empirically found to lie between 
-0.5 to -1 eV/atom; if the depths of the three edges of the ternary hull are in this range, stable and metastable ternary phases are found.
Based on this idea, we performed preliminary calculations of the binary convex hulls for the different combinations of elements in the diamond anvil cell.
This allowed us to focus our resources on the most promising ternary phases.

\paragraph*{Choice of the elements}
The elements that we chose to analyze are based on a few considerations on the experimental report. According to the authors, the first onset of a superconducting transition occurred at 294 K (not far from the reported value for LaH$_{10}$), and \tc{} gradually shifted towards higher temperatures, upon further heating. It seems reasonable to assume that LaH$_{10}$ was formed first, and, with the subsequent heating cycles, the sample underwent further structural transitions into an unknown multinary hydride phase, incorporating one or more of the elements present in the diamond anvil cell during the experiment.
The diamond anvil cell was loaded with pure lanthanum and ammonia borane (NH$_3$BH$_3$), which acts both as a hydrogen source and as pressure medium, and the authors mention as possible contaminants also platinum, gallium, and gold from the electrodes, and carbon from the epoxy binder. 
In principle, any combination of these elements may be responsible for the observed hot superconductivity phase.

\begin{table*}[ht]
	\begin{tabular}{cccccc}
		\hline\hline
		$X$ 	& Source  & \makecell{$\Delta H (La-X)$ \\ (eV/atom)} 	& \makecell{$\Delta H (X-H)$ \\(eV/atom)} & Stable	& Metastable$^{\dagger}$ 		 \\
		\hline
		\textbf{B} & H- source, gasket insert 		  &	-0.27			&	-0.05 	& 3		& 17	\\
		\textbf{N} & H- source, gasket insert  &	-1.65			&	-0.85 & 4 & 23 	\\
		\textbf{Pt} 	& Electrodes 	&	0.10		&	-0.30	& 0  & 0		 	\\
		\textbf{C} & Electrodes, epoxy &	0.13 		 &	-0.05	 	& 0	 & 0 	  	\\
		Au			& Electrodes 	&	0.15		&	0.05	 & -			& -				\\
		Ga			& Electrodes		&	0.10		&	-0.06	& -			& -			\\	
		\hline\hline
	\end{tabular}
	\caption{
		Elements included in the structural searches for a La-$X$-H phase. In the second column we report the main sources of each element in the experiment. $\Delta H (La-X)$ and $\Delta H (X-H)$ show the \textit{depth} of the La-$X$ and $X$-H binary convex hull for La-$X$-H system ($X$ = B, N, Pt, C, Ga, Au) in eV/atom. \textbf{Bold} indicates compositions for which the whole ternary convex hull was calculated. The depth is calculated considering the enthalpy difference between the energy of the lowest-lying composition on the binary hull, and its elemental components. A positive number means that no stable binary phase exist, and is the enthalpy of the lowest unstable binary phase observed. The depth of the La-H binary hull is -0.57 eV/atom. The fifth and sixth columns indicate the number of thermodynamically stable and metastable ternary phases. A structure is considered \textit{metastable} if it is within 50 meV/atom from the convex hull. }
	\label{tab:binary_depth}
\end{table*}
\paragraph*{Preliminary scan}
In Tab. \ref{tab:binary_depth} we report the \textit{depth} of the binary convex hulls together with the number of predicted stable ternary compositions. We predict no stable binary phases along the La-Ga, La-Au, Au-H, La-Pt, La-C lines, and we find very shallow minima for Ga-H, and C-H. On the other hand, we observe that the depth of La-N, N-H, La-B, and B-H binary hulls is in the optimal range.
Based on these indications, together with the scarcity of gallium and gold in the cell, we dismissed further investigation of La-Au-H and La-Ga-H, and we calculated the ternary convex hulls for La-N-H, La-B-H, La-Pt-H, and La-C-H, sampling 3000 structures for each diagram. This second preliminary search resulted in no stable (or weakly metastable) compositions for La-Pt-H and La-C-H, while several stable/metastable ones were already found for La-N-H, and La-B-H, supporting the empirical stability arguments based on binary hulls. This led us to dismiss further investigation of La-Pt-H and La-C-H, and to devote our resources to a better sampling of La-N-H and La-B-H. Boron and nitrogen were also the most abundant elements other than hydrogen and lanthanum during the experiment, and thus are the most likely candidates to form ternary La-$X$-H hydrides. The existence of high-pressure compounds involving La, N, and B, as well as metal boron hydrides also supports this hypothesis \cite{Sood_SSC_2004_LaB6, Uchida_Nature_1994_LaNiBN,DiCataldo_PRB_2020_CaBH,Kokail_PRM_2017_LiBH}. 

\paragraph*{Ternary Phase Diagram}
Having decided to focus on La-N-H and La-B-H, we improved our structural searches by performing a second structural search, sampling a total of 5000 structures, which add to the previously sampled 3000. In addition, we performed structural searches along several pseudo-binary phases, bringing the total number of structures sampled to about 12000.
In these refined searches, we sampled unit cells as large as 48 atoms.
See the Supplemental Material for further details. 

The convex hull diagrams for La-N-H and La-B-H are reported in Fig.~\ref{fig:hulls}, and the main results are summarized in table \ref{tab:binary_depth}.

\begin{figure*}[ht]
	\includegraphics[width=0.9\linewidth]{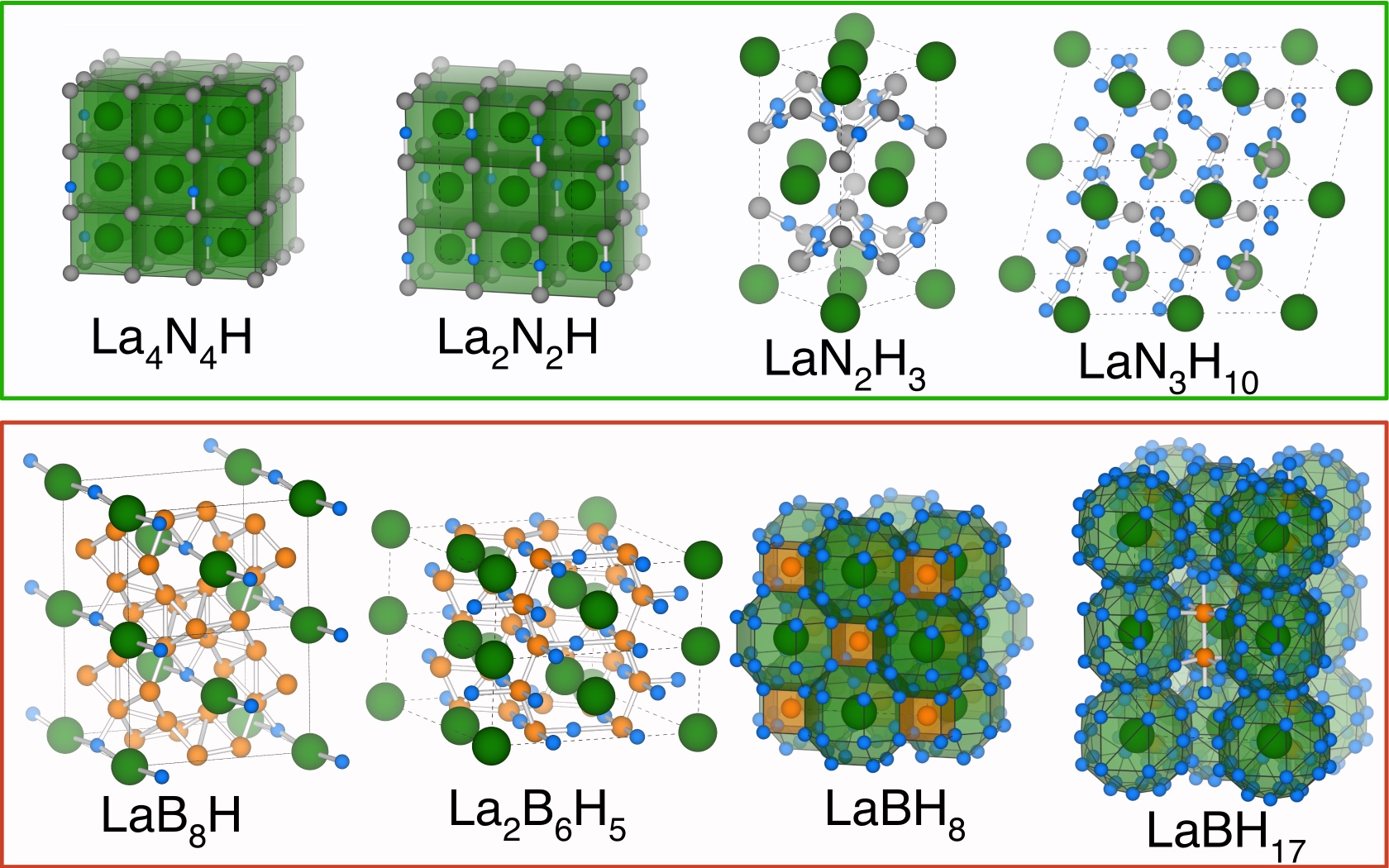}
		\caption{Crystal structures of the thermodynamically stable La-B-H and La-N-H phases. La, B, N and H atoms are shown as green,orange, gray, and blue spheres, respectively. Polyhedral surfaces match the color of the bonding atom.}
	\label{fig:all_structures}
\end{figure*}

The La-N-H convex hull contains four stable ternary phases: LaN$_2$H$_3$, LaN$_3$H$_{10}$, La$_2$N$_2$H, and La$_4$N$_4$H, as well as several metastable phases along the
(LaN)$_x$-H$_{1-x}$ line. We anticipate that none of the (meta)stable structures predicted in the La-N-H phase diagram is a likely candidate for HTCS since for most structures the hydrogen content is low, and H-rich weakly metastable structures are either low-symmetry or insulating. 

In the La-B-H phase diagram we find three stable intermediate compositions: LaBH$_{8}$, La$_2$B$_6$H$_5$, and LaB$_8$H. In addition, several H-rich phases are predicted to
be metastable (within 50 meV/atom from the hull), suggesting that there is a strong tendency for La and B to form H-rich phases. In this work we focused on the breadth of the search, rather than on the accuracy of the single phase diagrams, and 
restricted our search to unit cells with a maximum number of 48 atoms; hence,
we cannot rule out the possibility that other high-symmetry phases with high hydrogen content  compete with
our best structures. In some of searches, metastable phases with high H content were sampled, showing disordered lattices
and sometimes segregated H$_2$ molecules. This is a rather strong indication that high-symmetry phases with high hydrogen content
and complex stoichiometry may indeed lie on the hull, but we could not find them due to limited resolution. 

%%The presence of disorder and sometimes segregated H$_2$ molecules among these phases suggests that 
%%

The crystal structures of the stable La-N-H and La-B-H phases are shown in Fig.\ref{fig:all_structures}; additional information can be found in the Supplemental Material \footnote{The Supplemental Material is available at..}. 
The La$_4$N$_4$H and La$_2$N$_2$H phases are characterized by a La-N sublattice with a cubic CsCl arrangement, in which hydrogen occupies the interstitial sites, with a H-H distance (d$_{H-H}$) of 3.6 and 2.6 \AA, respectively. On the other hand, the LaN$_2$H$_3$ structure is characterized by the presence of La layers, alternated with a N-H network, and a H-H distance of 1.4 \AA, while LaN$_3$H$_{10}$ exhibits a disordered mixture of H$_2$ (d$_{H-H}$ = 0.74 \AA), NH, NH$_2$, and NH$_3$ molecules scattered around a La atom. The few metastable phases at high hydrogen content that we predict, are also characterized by the presence of disordered H$_2$ and NH$_x$ molecules.

The LaB$_8$H phase exhibits a dense B-B network around each La atom, identical to the one predict for LaB$_8$, with hydrogen occupying interstitial positions between the second-nearest La-La atoms, with a H-H distance of 3.7 \AA. The La$_2$B$_6$H$_5$ phase is characterized by the presence of two polymeric chains of BH$_2$ and B$_2$H$_2$, connected by a shared hydrogen atom, resulting in a H-H distance of 1.4 \AA, while the La atom acts as a spacer among the polymers. The LaBH$_{8}$ phase exhibits a densely-packed structure with two compenetrating face-centered cubic lattices for La and B, while H atoms form a rhombicuboctahedron around La and a cube around B. Nearest-neighboring hydrogen atoms form tetrahedra with a H-H distance of 1.33 \AA. This phase is particularly interesting at lower pressures, as it remains stable down to 40 GPa, where it exhibits high-\tc{} superconductivity -- further details have been discussed in Ref.~\onlinecite{DiCataldo_arXiv_LaBH8_2021}, while at 300 GPa the \tc{} is strongly suppressed by phonon hardening.

In addition to the thermodynamically stable phases, we chose to include in our pool of interesting structures also the metastable phase for LaBH$_{17}$, due to its structural similarity with sodalite-like hydrides, and its high hydrogen content. This phase most likely represents a member of a larger class of very H-rich structures, which we could not sample
completely due to limited cell size.
Its crystal structure presents a slightly distorted orthorhombic structure ($\alpha$-LaBH$_{17}$), in which a cage of 32 H atoms surrounds a La atom. Neighboring cages are alternated with B$_{2}$H$_{10}$ molecules. The cages are stacked along the vertical axis, sharing a slightly distorted hydrogen hexagon, with a H-H distance of 0.95 \AA. At 500 GPa this phase undergoes a structural transition into a tetragonal structure ($\beta$-LaBH$_{17}$), which is essentially degenerate in energy with the orthorhombic one and only differs for a small rotation of the H cages with respect to the stacking axis and the regularization of the hydrogen hexagon. Most likely, the inclusion of quantum effects would remove the distortion at lower pressure, in the same way as reported for other superhydrides \cite{Mauri_Nature_2016_SH3, Mauri_Nature_2020_LaH}. Both LaBH$_8$ and LaBH$_{17}$ have a relative hydrogen content above 50\%, and are characterized by a symmetric lattice, with lanthanum and boron atoms encaged into hydrogen polyhedra, which form a sponge-like lattice, and hence are an example of a ternary hydride with a crystal structure that is reminiscent of sodalite clathrate hydrides. The La-B-H system is the only one, among those examined, where structures of this type are realized. It is very likely that other structures with even higher H content
may appear on the hull; identifying them would require focused sampling of specific H-rich compositions, with
very large unit cell, which is a formidable task for current crystal structural search methods.

\begin{table*}[!ht]
	\begin{tabular}{ccccccccccc}
		\hline\hline
		Composition 	& Space Group 	& P 	& $\Delta$H & d$_{H-H}$ & N(E$_F$) 						& N$_H$/N($E_{F}$) & $\lambda$ &$\omega_{log}$   & \tc{}$^{ME}$ &\\
		&			& (GPa)	& (eV/atom) &	 (\AA) 		& (10$^{3}$spin$^{-1}$eV$^{-1}$\AA$^{-3}$)  &							&					&		(K)		&		(K) 	&\\
		%LaB$_2$H$_6$		& 227/227		&150&					& 1.28	 		&0.55		&15\%	&0.75		&734			& 25	&\\
		%LaB$_2$H$_6$		& 227/227		&300 &			 		& 1.14			&	0.33	&14\%	&0.48		&754			& 5 &\\
		La$_4$N$_4$H	&	38					& 300&	-0.09					&	3.6		& 16.9		&	1\% &	0.24	&	434		& 0	& \\
		La$_2$N$_2$H	&	63					&300 &  -0.15				& 2.6		& 4.2		& 3\%	&	0.15	&	719	&	0	& \\
		LaN$_2$H$_3$	&	66					&300&	-0.17				& 1.4		& 10.3		& 1\%		& 0.33	&	966	&	1	&	\\
		LaN$_3$H$_{10}$  & 1					& 300 &	 -0.25				 &  0.74	&	-		 &	-		 &		-		&				&	-	& \\
		LaB$_8$H			& 5						&300& -0.08				&  	3.7			& 8.3		&  4\%	 &	0.44		&	973		& 8	& \\
		La$_2$B$_6$H$_5$ &	8				&300 &	-0.04			& 1.5			& 12.0		& 21\%	& 0.47			&	998		&6	&	\\
		%LaBH$_8$			& 225/225 		& 50 &		-0.05			&	1.65		&	7.0		&73\%	&1.99		&736			& 128 &\\
		%LaBH$_8$				& 225/225 		& 100 &		 			&	1.53		&	0.27	&71\%	&1.03		&	1233		& 104	&\\
		%LaBH$_8$				& 225/225 		& 150  &				& 1.46			&	0.25	&67\%	&0.73		&1440			& 46	&\\
		LaBH$_8$				& 225 		& 300  & -0.24			&	1.33		&	7.4		&62\%	&0.53		&	1731		& 14	& \\
		$\alpha$-LaBH$_{17}$& 23 		&300  &	-0.12	&	0.95		& 7.8		&63\%		&3.3		&	414			& 180	& \\
		$\beta$-LaBH$_{17}$	& 97$^{*}$&300 & -0.12		&	0.96	  & 7.9		&64\%		&2.3$^{**}$	&	759		& 179	& \\
		LaH$_{10}$				& 225		&300& -				&	1.06-1.14      	&	16.4	&  62\%		&1.9			& 1575		& 249	&\\
		\hline
		\hline\hline
	\end{tabular}
	\caption{
		Electronic and superconducting properties of selected ternary phases of La-N-H and La-B-H. The first column shows the composition, the second column indicates the space group. The fourth column $\Delta$H indicates the energy relative to a decomposition into the lowest-energy binary phases. d$_{H-H}$ indicates the average H-H distance in \AA{}ngstrom. In the sixth and seventh column the electronic DOS at the Fermi level N(E$_F$) and its relative hydrogen character are reported. The DOS is shown per unit volume to allow for comparison between different pressures. The electron-phonon coupling coefficient $\lambda$ and the average phonon frequency $\omega_{log}$ are defined in the Supplemental Material. The superconducting critical temperature  \tc{}$^{ME}$ was calculated by solving the isotropic Migdal-Eliashberg equations -- for details see the Supplemental Material. $*$ the structure is dynamically unstable near the M point. $**$ cutting imaginary frequencies. }
	\label{tab:electronic_structure}
\end{table*}

\section{Electronic Properties}
\begin{figure*}[ht]
	\centering
	\includegraphics[width=0.90\linewidth]{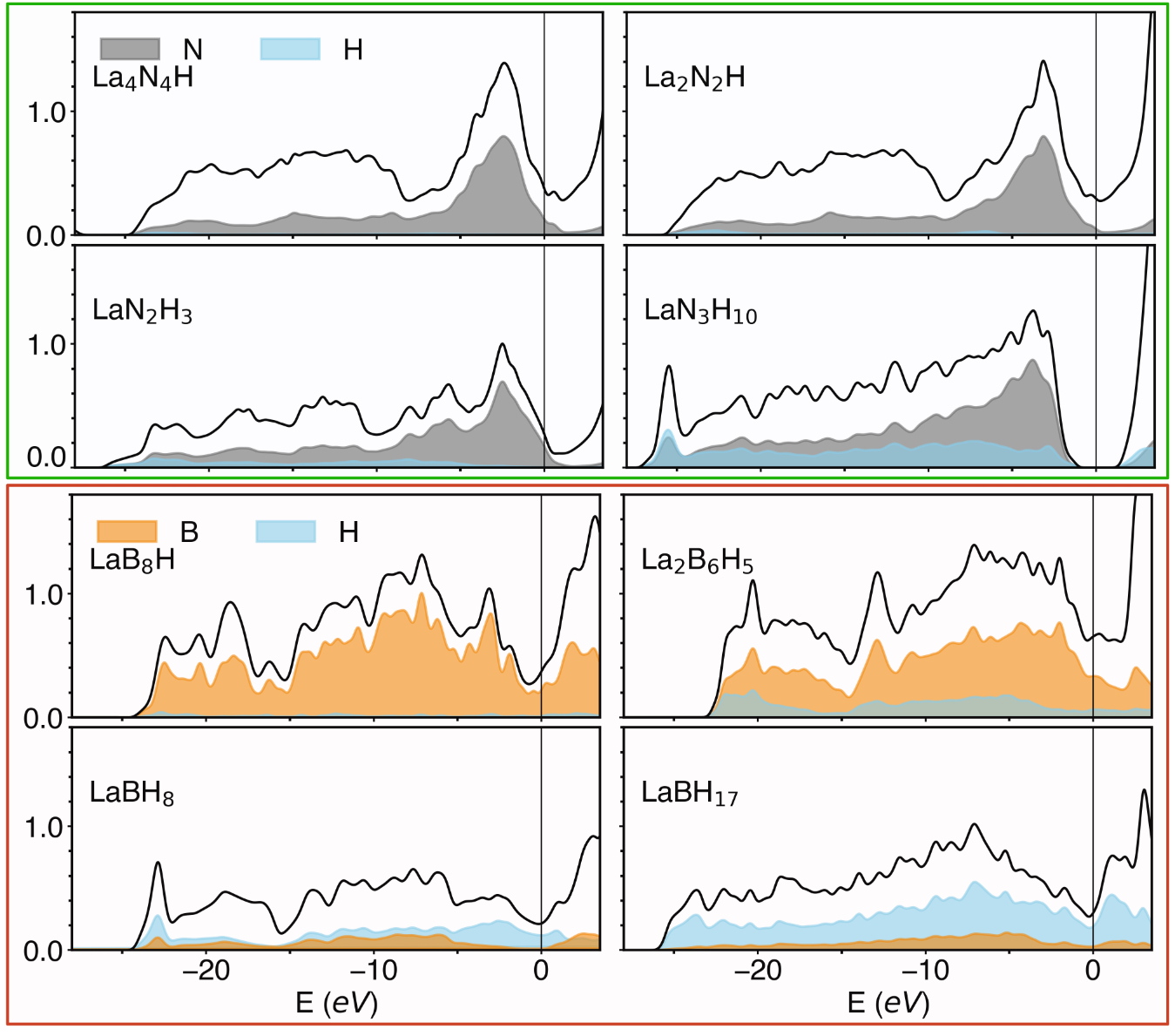}
	\caption{Total and atom-projected DOS in units of $spin^{-1}$$eV^{-1}$ for stable La-N-H and La-B-H phases at 300 GPa. The total DOS and its projection onto La, N, B, and H are shown as black lines, and green, gray, orange, and blue filled lines, respectively. The Fermi energy is set as the zero.}
	\label{fig:ternariesdosall}
\end{figure*}

In Fig. \ref{fig:ternariesdosall} we report the total and atom-projected electronic Density of States (DOS) for the stable La-N-H and La-B-H phases. The structures for La$_4$N$_4$H are La$_2$N$_2$H are metallic and the partial DOS in the valence region is dominated by lanthanum and nitrogen, with little contribution from interstitial hydrogen. In LaN$_2$H$_3$ hydrogen gives a rather small contribution to the DOS from -25 to -5, while nitrogen strongly contributes in the -10 to 0 eV range, and makes up most of the states at the Fermi level. LaN$_3$H$_{10}$, on the other hand, is characterized by an insulating structure, with a band gap of 2.4 eV. The states in the valence band have a predominant nitrogen and hydrogen character. The two DOS's follow each other rather closely, indicating significant covalent N-H bonding in the N-H$_x$ molecules. Overall, none of the La-N-H structures shown exhibit the typical characteristics of superconducting hydrides: high-symmetry structures, with high hydrogen content, and a large DOS at the Fermi level which is mostly derived from hydrogen. In La-N-H, we observe that at low concentration, hydrogen plays no significant role in the band structure near the Fermi energy, while at higher concentration, the structures formed are low-symmetry molecular crystals, either insulating, or with negligible contribution of hydrogen to the states near the Fermi energy.

The stable La-B-H structures (and metastable LaBH$_{17}$), on the other hand, are all metallic. In particular, LaB$_8$H is characterized by a partial DOS with a predominant boron contribution, while hydrogen contributes only very little.  The La$_2$B$_6$H$_5$ structure exhibits a much larger, and rather constant, contribution of lanthanum to filled states. In both cases, however, hydrogen does not contribute significantly to the states at the Fermi level, nor does it play a significant role in the electronic structure. The structure for LaBH$_{8}$ is characterized by a partial DOS character rather equally distributed among La, B, and H in the -25 to -5 eV range, which gives way to a hydrogen-dominated DOS in the -5 to 1 eV range, including states at the Fermi level. Last, LaBH$_{17}$ exhibits a strong hydrogen character at all energies, as the structure is characterized by a weakly covalent hydrogen network, and a very high hydrogen concentration. LaBH$_{8}$ and LaBH$_{17}$ are the only compositions for which the electronic structures are similar to the ones reported for sodalite-like superhydrides, i.e. a large hydrogen fraction of the states at the Fermi level N$_H$/N(E$_F$), as well as a large overall Density of States at the Fermi level (N(E$_F$)). Moreover, the structures of both LaBH$_8$ and LaBH$_{17}$ also satisfy a geometrical prerequisite: it was observed that the presence of weakly-covalent hydrogen-hydrogen bonds positively correlates with \tc{}\cite{Errea_arXiv_2021_hydridebased}. These bonds are associated with a H-H distance larger than 0.80 \AA{} (i.e. no H$_2$ molecules), but smaller than 2 \AA{} (i.e. significant H-H interaction). When the H-H interatomic distance satisfies these constraints the H lattice is so dense that a quasi-free electron gas is realized \cite{Borinaga_PRB_2016_atomicH, Needs_PRL_2014_atomicH}. A possible descriptor to characterize the nature of the H-H bond in hydrides, called \textit{connectivity}, was recently proposed by Belli et al. \cite{Errea_arXiv_2021_hydridebased}. The connectivity value $\phi$ represents the highest value of the Electron Localization Function (ELF) for which the isocontour spans the cell without discontinuities in all directions. In particular, isolated high ELF regions surrounding H-H bonds  are associated with \textit{molecular} hydrogen, while a $\phi$ between approximately 0.4 and 0.8 is considered an indication of weak covalent bonds. For LaBH$_8$ and LaBH$_{17}$ at 300 GPa we find a connectivity value $\phi$ of 0.33 and, 0.43, respectively and no molecular bonds, which classifies LaBH$_{17}$, and possibly also LaBH$_{8}$, as a weak-covalent hydride. The ELF corresponding to the $\phi$ value for the two structures is shown in Fig. S1.

\section{Superconducting Properties}
In order to asses the superconductiving properties of the predicted La-B-H and La-N-H structures, we calculated their vibrational and superconducting properties. %
Phonon frequencies were computed at the harmonic level, and the superconducting  \tc{} 
due to $e-ph$ interaction was calculated by numerically solving the T-dependent isotropic Migdal-Eliashberg equations, using a standard value of $\mu^{*}$ of 0.10~\cite{Allen_PRB_1975_McMillan, Carbotte_RevModPhys_1990_sc}. Our estimates do not take into account anharmonicity, whose effect should be to make our structures stable at lower pressures than the nominal harmonic instability pressure \cite{Heil_PRB_2019, Mauri_Nature_2016_SH3, Eremets_Nature_2019_LaH}. 
A summary of the electronic and vibrational properties of the stable phases of La-B-H and La-N-H at 300 GPa is reported in Tab.\ref{tab:electronic_structure}. All structures are dynamically stable, and the predicted \tc's are 8 K in LaB$_8$H, 6 K in La$_2$B$_6$H$_5$, 14 K in LaBH$_8$ and 170 K in LaBH$_{17}$, while La$_4$N$_4$H, La$_2$N$_2$H, and LaN$_2$H$_3$ exhibit a \tc{} below 1 K.
\begin{figure*}[ht]
	\centering
	\includegraphics[width=0.90\linewidth]{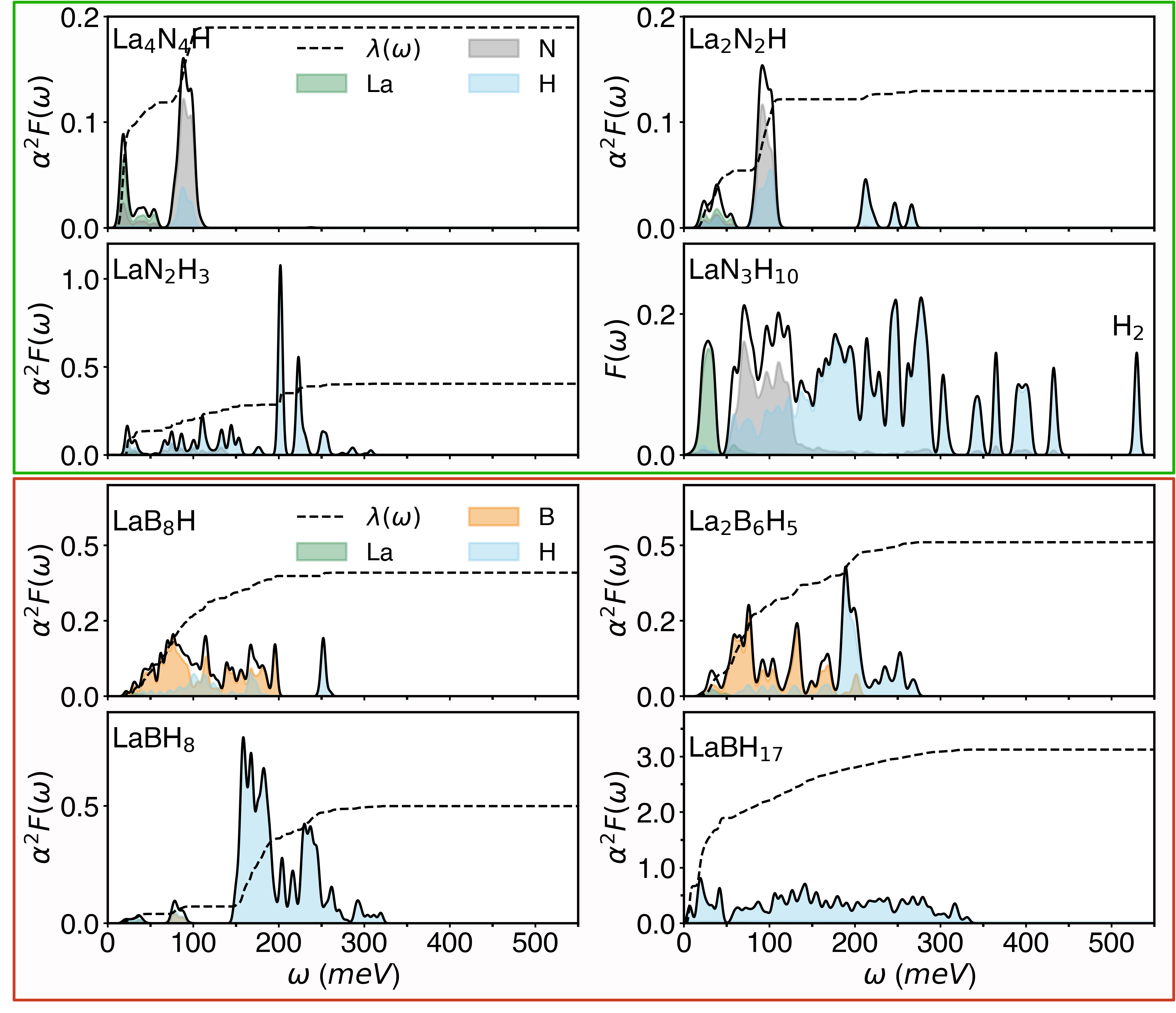}
	\caption{Total and atom-projected Eliashberg function [$\alpha^2 F(\omega)$, solid lines], and $\omega$-dependent \ep coupling [$\lambda(\omega)$, dashed lines] for stable La-B-H and La-N-H hydrides. For the insulating LaN$_3$H$_{10}$ we report the total and atom-projected phonon DOS  [$F(\omega)$, solid lines]. The atom projections on La, B, N, and H are shown in green, gray, orange, and blue, respectively. The Eliashberg function and $\omega$-dependent \ep coupling $\lambda(\omega)$ are defined in the Supplemental Material. \textbf{Note:} due to the large differences in values, the $y$-axis scale is different for each subfigure.}
	\label{fig:a2f}
\end{figure*}

In Fig. \ref{fig:a2f} we report the total and the atom-projected Eliashberg functions for all stable structures (except LaN$_{3}$H$_{10}$, for which we report the phonon DOS, as it is insulating), as well as for metastable LaBH$_{17}$. As shown in the figure, in both La$_4$N$_4$H and La$_2$N$_2$H it is mostly nitrogen vibrations which contribute to the overall \ep{} coupling, which is extremely small. In La$_4$N$_4$H, the H-derived modes with frequencies above 150 meV are absent, while they are present, albeit very weakly coupled, in La$_2$N$_2$H. It is clear that in these two structures hydrogen, which occupies the interstitial sites, does not play a significant role in the bonding or in the properties, and thus cannot contribute to the coupling. The spectrum of LaN$_2$H$_3$ is characterized by a few narrow peaks between 100 and 250 meV, with strong hydrogen character. It is interesting to compare these results with the phonon spectrum of LaN$_3$H$_{10}$ which, unlike the others, exhibits a peak around 530 meV, which arises from the H$_{2}$ molecule vibron. In these two structures we have shown that the structural and electronic properties are dictated by nitrogen, and is again not particularly surprising not to find high-\tc{} superconductivity.
In LaB$_{8}$H the vibrational spectrum is dominated by boron, while in La$_2$B$_6$H$_5$ about 30\% of it is hydrogen. In both cases however, the integrated \ep{} coupling $\lambda$ is around 0.5, i.e. too low to lead to an appreciable \tc{}. In these cases the hydrogen content is rather low, therefore it is the B-B covalent bonding that dominates.

LaBH$_{8}$ and LaBH$_{17}$ are significantly different: here essentially all of the coupling is concentrated into hydrogen modes. In particular, in LaBH$_{8}$ the coupling largely comes from modes around 170 meV, yielding a high $\omega_{log}$ of 135 meV, and a relatively low $\lambda=0.5$. LaBH$_{17}$, on the other hand, exhibits a large value of the Eliashberg function at all energy ranges,
concentrated on hydrogen modes. At 300 GPa, $\alpha$-LaBH$_{17}$ is on the verge of a structural instability, as witnessed by its small \omlog, and large electron-phonon coupling constant $\lambda = 3.3$. At higher pressures the phonon harden.
Both LaBH$_8$ and LaBH$_{17}$ exhibit the typical characteristic of high-\tc{} superhydrides, i.e. the presence of an interconnected, metallic hydrogen sublattice, which is manifested both in the large fraction of hydrogen states in the DOS at the Fermi level, and in a rather uniform distribution of the electron-phonon cuopling over all phonon modes, as observed in 
binary high-\tc{} sodalite-like hydrides.
At 300 GPa, only LaBH$_{17}$ is a high-\tc{} superconductor.  
In another publication, we studied the superconducting behavior of LaBH$_8$ as a function of pressure in greater detail \cite{DiCataldo_arXiv_LaBH8_2021}. This structure remains dynamically stable down to 40 GPa, where the phonon modes are softer, and the \tc{} reaches 126 K. At 300 GPa however, the pressure-induced hardening of the phonon modes is so strong that suppresses the high-\tc{}. 
All other structures are undoubtedly not superhydrides, and given their small hydrogen fraction it is not surprising that they are not high-\tc{} superconductors.

To rule out that the discrepancy between predicted \tc{} and experiments may be due to the pressure shift introduced to 
empirically take into account the effect of quantum lattice fluctuations we recomputed the critical temperatures of LaBH$_8$ and LaBH$_{17}$ also at
150 GPa, i.e. below the maximum pressure reported in experiments. Here, LaBH$_8$ is dynamically stable, with a \tc{} of 40 K, while LaBH$_{17}$ has
a weak dynamical instability; neglecting imaginary frequencies, the predicted \tc{} is 223 K, i.e. well below 550 K.

We can also estimate, albeit in a rather qualitative fashion, the effect of possible mechanisms that may positively influence the \tc{}
of related ternary and multinary La-B-H phases. 
An obvious observation is that the Fermi energy in LaBH$_{17}$ lies exactly in correspondence of a pseudogap. Carbon substitution at the boron site could sensibly enhance the \tc{} through charge doping; for example, a 50\% replacement of boron with carbon would increase the DOS at the Fermi level by approximately a factor of two, and boost the \tc{} to about 290 K. If we assume that the Eliashberg function is rigidly multiplied by a factor equal to the increase in the DOS, in order to achieve 550 K one would need approximately an eight times larger DOS in LaBH$_{8}$, and a four times larger DOS at the Fermi level in LaBH$_{17}$. At most, a 100\% substitution of boron with carbon would shift the Fermi energy enough to boost the DOS by a factor of two and a half, and the \tc{} to about 310 K, i.e. above room temperature, but well below 550 K.
Another possibility is to consider phases with a higher H content than LaBH$_{17}$; the shape of the La-B-H ternary hull gave strong indications
that they may form. In this case, one may speculate that average phonon frequencies may be increased compared to LaBH$_{17}$, leading to an
effective boost in \tc{}. However, even a doubling of all phonon frequencies of LaBH$_{17}$, which is extremely unlikely, would be sufficient
to bring the \tc{} {\em only} to 360 K.

\section{Conclusions}
In conclusion, following a recent experimental report of \textit{hot} superconductivity at 550 K in a material with undetermined composition and structure \cite{superconductivity_550K}, we investigated from first-principles the high-pressure phase diagram of the most likely combinations of elements which could have formed, i.e. La-$X$-H ternary hydrides ($X$ = B, N, Pt, Au, Ga, C), looking for a candidate to explain the experimental results. The choice of La-based ternary hydrides is motivated by the \tc{} measured after the first heating cycle, which is compatible with that of LaH$_{10}$. As $X$ element we considered all the elements that were reported to be present in the diamond anvil cell during the experiment: boron, nitrogen, hydrogen, and traces of platinum, gold, gallium and carbon. 

In order to evaluate which ternary hydrides are thermodynamically favorable, we used variable-composition, evolutionary algorithms, at increasing levels of accuracy.
According to our calculations, only La-N-H and La-B-H can form stable ternary phases, and only in La-B-H  do we  predict the formation of H-rich, highly symmetric structures, which can host high-\tc{} superconductivity. In particular, we identified a high-\tc{} tetragonal LaBH$_{17}$ phase, characterized by a dense hydrogen sublattice which is reminiscent of other high-\tc{} binary sodalite-like hydrides. For this structure we predicted a \tc{} of 180 K at 300 GPa by numerically solving the isotropic Migdal-Eliashberg equations. This result is way too far from the reported value of \textit{hot} superconductivity to attribute the difference to numerical errors, and even within the most optimistic doping scenarios we could not increase \tc{} above 360 K.
The discrepancy is too large for anharmonic lattice effects to affect our main conclusions.

While none of the binary or ternary phases of the elements considered in this work can explain the extreme \tc{}s reported,
the La-B-H system
represents a very interesting starting point for further superconductivity studies.
In fact, the extreme complexity of a ternary search limited our calculations in the maximum size of the unit cell, and the extent of the sampling for each composition, but our calculations and stability arguments indicate that the formation of high-H content La-B-H phases, or even quaternary La-B-N-H phases, with high \tc{} is extremely likely.
Therefore, we urge the authors of Ref.~\onlinecite{superconductivity_550K} to repeat their experiments under controlled conditions; it would be interesting, for example, to repeat the experiments employing diborane (B$_2$H$_6$), instead of ammonia borane as a hydrogen source, to discriminate between purely ternary La-B-H phases and quaternary La-B-N-H ones.

We hope that a more precise determination of the critical temperature and a clearer indication of the elements and crystal structures will help
elucidate the fascinating high-pressure physics of these systems.
\section{Acknowledgements}
We thank Antonio Sanna for kindly sharing with us the code for solving the isotropic Migdal-Eliashberg equations. The authors acknowledge computational resources from the dCluster of the Graz University of Technology and the VSC3 of the Vienna University of Technology, and support through the FWF, Austrian Science Fund, Project P30269-N36 (Superhydra). L. B. acknowledges funding through Progetto Ateneo Sapienza 2017-18-19 and computational Resources from CINECA, proj. Hi- TSEPH.
%\bibliographystyle{apsrev4-1}
%\bibliography{ternaries_bib}

\begin{thebibliography}{36}%
	\makeatletter
	\providecommand \@ifxundefined [1]{%
		\@ifx{#1\undefined}
	}%
	\providecommand \@ifnum [1]{%
		\ifnum #1\expandafter \@firstoftwo
		\else \expandafter \@secondoftwo
		\fi
	}%
	\providecommand \@ifx [1]{%
		\ifx #1\expandafter \@firstoftwo
		\else \expandafter \@secondoftwo
		\fi
	}%
	\providecommand \natexlab [1]{#1}%
	\providecommand \enquote  [1]{``#1''}%
	\providecommand \bibnamefont  [1]{#1}%
	\providecommand \bibfnamefont [1]{#1}%
	\providecommand \citenamefont [1]{#1}%
	\providecommand \href@noop [0]{\@secondoftwo}%
	\providecommand \href [0]{\begingroup \@sanitize@url \@href}%
	\providecommand \@href[1]{\@@startlink{#1}\@@href}%
	\providecommand \@@href[1]{\endgroup#1\@@endlink}%
	\providecommand \@sanitize@url [0]{\catcode `\\12\catcode `\$12\catcode
		`\&12\catcode `\#12\catcode `\^12\catcode `\_12\catcode `\%12\relax}%
	\providecommand \@@startlink[1]{}%
	\providecommand \@@endlink[0]{}%
	\providecommand \url  [0]{\begingroup\@sanitize@url \@url }%
	\providecommand \@url [1]{\endgroup\@href {#1}{\urlprefix }}%
	\providecommand \urlprefix  [0]{URL }%
	\providecommand \Eprint [0]{\href }%
	\providecommand \doibase [0]{http://dx.doi.org/}%
	\providecommand \selectlanguage [0]{\@gobble}%
	\providecommand \bibinfo  [0]{\@secondoftwo}%
	\providecommand \bibfield  [0]{\@secondoftwo}%
	\providecommand \translation [1]{[#1]}%
	\providecommand \BibitemOpen [0]{}%
	\providecommand \bibitemStop [0]{}%
	\providecommand \bibitemNoStop [0]{.\EOS\space}%
	\providecommand \EOS [0]{\spacefactor3000\relax}%
	\providecommand \BibitemShut  [1]{\csname bibitem#1\endcsname}%
	\let\auto@bib@innerbib\@empty
	%</preamble>
	\bibitem [{\citenamefont {Drodzov}\ \emph
		{et~al.}(2015{\natexlab{a}})\citenamefont {Drodzov}, \citenamefont {Eremets},
		\citenamefont {Troyan}, \citenamefont {Ksenofontov},\ and\ \citenamefont
		{Shylin}}]{Eremets_Nature_2015_SH3}%
	\BibitemOpen
	\bibfield  {author} {\bibinfo {author} {\bibfnamefont {A.~P.}\ \bibnamefont
			{Drodzov}}, \bibinfo {author} {\bibfnamefont {M.~I.}\ \bibnamefont
			{Eremets}}, \bibinfo {author} {\bibfnamefont {I.~A.}\ \bibnamefont {Troyan}},
		\bibinfo {author} {\bibfnamefont {V.}~\bibnamefont {Ksenofontov}}, \ and\
		\bibinfo {author} {\bibfnamefont {S.~I.}\ \bibnamefont {Shylin}},\
	}\href@noop {} {\bibfield  {journal} {\bibinfo  {journal} {Nature}\ }\textbf
		{\bibinfo {volume} {525}},\ \bibinfo {pages} {73} (\bibinfo {year}
		{2015}{\natexlab{a}})}\BibitemShut {NoStop}%
	\bibitem [{\citenamefont {Duan}\ \emph {et~al.}(2014)\citenamefont {Duan},
		\citenamefont {Liu}, \citenamefont {Tian}, \citenamefont {Li}, \citenamefont
		{Huang}, \citenamefont {Zhao}, \citenamefont {Yu}, \citenamefont {Liu},
		\citenamefont {Tian},\ and\ \citenamefont {Cui}}]{Duan_SciRep_2014_SH}%
	\BibitemOpen
	\bibfield  {author} {\bibinfo {author} {\bibfnamefont {D.}~\bibnamefont
			{Duan}}, \bibinfo {author} {\bibfnamefont {Y.}~\bibnamefont {Liu}}, \bibinfo
		{author} {\bibfnamefont {F.}~\bibnamefont {Tian}}, \bibinfo {author}
		{\bibfnamefont {D.}~\bibnamefont {Li}}, \bibinfo {author} {\bibfnamefont
			{X.}~\bibnamefont {Huang}}, \bibinfo {author} {\bibfnamefont
			{Z.}~\bibnamefont {Zhao}}, \bibinfo {author} {\bibfnamefont {H.}~\bibnamefont
			{Yu}}, \bibinfo {author} {\bibfnamefont {B.}~\bibnamefont {Liu}}, \bibinfo
		{author} {\bibfnamefont {W.}~\bibnamefont {Tian}}, \ and\ \bibinfo {author}
		{\bibfnamefont {T.}~\bibnamefont {Cui}},\ }\href@noop {} {\bibfield
		{journal} {\bibinfo  {journal} {Scientific Reports}\ }\textbf {\bibinfo
			{volume} {4}},\ \bibinfo {pages} {6968} (\bibinfo {year} {2014})}\BibitemShut
	{NoStop}%
	\bibitem [{\citenamefont {Einaga}\ \emph {et~al.}(2016)\citenamefont {Einaga},
		\citenamefont {Sakata}, \citenamefont {Ishikawa}, \citenamefont {Shimizu},
		\citenamefont {Eremets}, \citenamefont {Drodzov}, \citenamefont {Troyan},
		\citenamefont {Hirao},\ and\ \citenamefont
		{Ohishi}}]{Eremets_NatPhys_2016_SH3}%
	\BibitemOpen
	\bibfield  {author} {\bibinfo {author} {\bibfnamefont {M.}~\bibnamefont
			{Einaga}}, \bibinfo {author} {\bibfnamefont {M.}~\bibnamefont {Sakata}},
		\bibinfo {author} {\bibfnamefont {T.}~\bibnamefont {Ishikawa}}, \bibinfo
		{author} {\bibfnamefont {K.}~\bibnamefont {Shimizu}}, \bibinfo {author}
		{\bibfnamefont {M.}~\bibnamefont {Eremets}}, \bibinfo {author} {\bibfnamefont
			{A.~P.}\ \bibnamefont {Drodzov}}, \bibinfo {author} {\bibfnamefont {I.~A.}\
			\bibnamefont {Troyan}}, \bibinfo {author} {\bibfnamefont {N.}~\bibnamefont
			{Hirao}}, \ and\ \bibinfo {author} {\bibfnamefont {Y.}~\bibnamefont
			{Ohishi}},\ }\href@noop {} {\bibfield  {journal} {\bibinfo  {journal} {Nature
				Physics}\ }\textbf {\bibinfo {volume} {12}},\ \bibinfo {pages} {835}
		(\bibinfo {year} {2016})}\BibitemShut {NoStop}%
	\bibitem [{\citenamefont {Zurek}\ and\ \citenamefont
		{Bi}(2019)}]{Zurek_JCP_2019_review}%
	\BibitemOpen
	\bibfield  {author} {\bibinfo {author} {\bibfnamefont {E.}~\bibnamefont
			{Zurek}}\ and\ \bibinfo {author} {\bibfnamefont {T.}~\bibnamefont {Bi}},\
	}\href@noop {} {\bibfield  {journal} {\bibinfo  {journal} {J. Chem. Phys.}\
		}\textbf {\bibinfo {volume} {150}},\ \bibinfo {pages} {050901} (\bibinfo
		{year} {2019})}\BibitemShut {NoStop}%
	\bibitem [{\citenamefont {Flores-Livas}\ \emph {et~al.}(2020)\citenamefont
		{Flores-Livas}, \citenamefont {Boeri}, \citenamefont {Sanna}, \citenamefont
		{Profeta}, \citenamefont {Arita},\ and\ \citenamefont
		{Eremets}}]{Boeri_PhysRep_2020_review}%
	\BibitemOpen
	\bibfield  {author} {\bibinfo {author} {\bibfnamefont {J.~A.}\ \bibnamefont
			{Flores-Livas}}, \bibinfo {author} {\bibfnamefont {L.}~\bibnamefont {Boeri}},
		\bibinfo {author} {\bibfnamefont {A.}~\bibnamefont {Sanna}}, \bibinfo
		{author} {\bibfnamefont {G.}~\bibnamefont {Profeta}}, \bibinfo {author}
		{\bibfnamefont {R.}~\bibnamefont {Arita}}, \ and\ \bibinfo {author}
		{\bibfnamefont {M.}~\bibnamefont {Eremets}},\ }\href@noop {} {\bibfield
		{journal} {\bibinfo  {journal} {Physics Reports}\ }\textbf {\bibinfo {volume}
			{856}},\ \bibinfo {pages} {1} (\bibinfo {year} {2020})}\BibitemShut {NoStop}%
	\bibitem [{\citenamefont {Semenok}\ \emph
		{et~al.}(2020{\natexlab{a}})\citenamefont {Semenok}, \citenamefont {Kruglov},
		\citenamefont {Savkin}, \citenamefont {Kvashin},\ and\ \citenamefont
		{Oganov}}]{Oganov_SSM_2020_MH}%
	\BibitemOpen
	\bibfield  {author} {\bibinfo {author} {\bibfnamefont {D.~V.}\ \bibnamefont
			{Semenok}}, \bibinfo {author} {\bibfnamefont {I.~A.}\ \bibnamefont
			{Kruglov}}, \bibinfo {author} {\bibfnamefont {I.~A.}\ \bibnamefont {Savkin}},
		\bibinfo {author} {\bibfnamefont {A.~G.}\ \bibnamefont {Kvashin}}, \ and\
		\bibinfo {author} {\bibfnamefont {A.~R.}\ \bibnamefont {Oganov}},\
	}\href@noop {} {\bibfield  {journal} {\bibinfo  {journal} {Current Opinion in
				Solid State and Materials Science}\ }\textbf {\bibinfo {volume} {24}},\
		\bibinfo {pages} {100808} (\bibinfo {year} {2020}{\natexlab{a}})}\BibitemShut
	{NoStop}%
	\bibitem [{\citenamefont {Drodzov}\ \emph {et~al.}(2019)\citenamefont
		{Drodzov}, \citenamefont {Kong}, \citenamefont {Besedin}, \citenamefont
		{Kuzonikov}, \citenamefont {Mozaffari}, \citenamefont {Balicas},
		\citenamefont {Balakirev}, \citenamefont {Graf}, \citenamefont {Prakapenka},
		\citenamefont {Greenberg}, \citenamefont {Knyazev}, \citenamefont {Tkacz},\
		and\ \citenamefont {Eremets}}]{Eremets_Nature_2019_LaH}%
	\BibitemOpen
	\bibfield  {author} {\bibinfo {author} {\bibfnamefont {A.~P.}\ \bibnamefont
			{Drodzov}}, \bibinfo {author} {\bibfnamefont {P.~P.}\ \bibnamefont {Kong}},
		\bibinfo {author} {\bibfnamefont {S.~P.}\ \bibnamefont {Besedin}}, \bibinfo
		{author} {\bibfnamefont {M.~A.}\ \bibnamefont {Kuzonikov}}, \bibinfo {author}
		{\bibfnamefont {S.}~\bibnamefont {Mozaffari}}, \bibinfo {author}
		{\bibfnamefont {L.}~\bibnamefont {Balicas}}, \bibinfo {author} {\bibfnamefont
			{F.~F.}\ \bibnamefont {Balakirev}}, \bibinfo {author} {\bibfnamefont {D.~E.}\
			\bibnamefont {Graf}}, \bibinfo {author} {\bibfnamefont {V.~B.}\ \bibnamefont
			{Prakapenka}}, \bibinfo {author} {\bibfnamefont {E.}~\bibnamefont
			{Greenberg}}, \bibinfo {author} {\bibfnamefont {D.~A.}\ \bibnamefont
			{Knyazev}}, \bibinfo {author} {\bibfnamefont {M.}~\bibnamefont {Tkacz}}, \
		and\ \bibinfo {author} {\bibfnamefont {M.~I.}\ \bibnamefont {Eremets}},\
	}\href@noop {} {\bibfield  {journal} {\bibinfo  {journal} {Nature}\ }\textbf
		{\bibinfo {volume} {569}},\ \bibinfo {pages} {528} (\bibinfo {year}
		{2019})}\BibitemShut {NoStop}%
	\bibitem [{\citenamefont {Somayazulu}\ \emph {et~al.}(2019)\citenamefont
		{Somayazulu}, \citenamefont {Ahart}, \citenamefont {Mishra}, \citenamefont
		{Geballe}, \citenamefont {Baldini}, \citenamefont {Meng}, \citenamefont
		{Struzhkin},\ and\ \citenamefont {Hemley}}]{Hemley_PRL_2019_LaH}%
	\BibitemOpen
	\bibfield  {author} {\bibinfo {author} {\bibfnamefont {M.}~\bibnamefont
			{Somayazulu}}, \bibinfo {author} {\bibfnamefont {M.}~\bibnamefont {Ahart}},
		\bibinfo {author} {\bibfnamefont {A.~K.}\ \bibnamefont {Mishra}}, \bibinfo
		{author} {\bibfnamefont {Z.~M.}\ \bibnamefont {Geballe}}, \bibinfo {author}
		{\bibfnamefont {M.}~\bibnamefont {Baldini}}, \bibinfo {author} {\bibfnamefont
			{Y.}~\bibnamefont {Meng}}, \bibinfo {author} {\bibfnamefont {V.~V.}\
			\bibnamefont {Struzhkin}}, \ and\ \bibinfo {author} {\bibfnamefont {R.~J.}\
			\bibnamefont {Hemley}},\ }\href@noop {} {\bibfield  {journal} {\bibinfo
			{journal} {Phys. Rev. Lett.}\ }\textbf {\bibinfo {volume} {122}},\ \bibinfo
		{pages} {027001} (\bibinfo {year} {2019})}\BibitemShut {NoStop}%
	\bibitem [{\citenamefont {Liu}\ \emph {et~al.}(2017)\citenamefont {Liu},
		\citenamefont {Naumov}, \citenamefont {Hoffmann}, \citenamefont {Ashcroft},\
		and\ \citenamefont {Hemley}}]{Ashcroft_PNAS_2017_LaH}%
	\BibitemOpen
	\bibfield  {author} {\bibinfo {author} {\bibfnamefont {H.}~\bibnamefont
			{Liu}}, \bibinfo {author} {\bibfnamefont {I.~I.}\ \bibnamefont {Naumov}},
		\bibinfo {author} {\bibfnamefont {R.}~\bibnamefont {Hoffmann}}, \bibinfo
		{author} {\bibfnamefont {N.~W.}\ \bibnamefont {Ashcroft}}, \ and\ \bibinfo
		{author} {\bibfnamefont {R.~J.}\ \bibnamefont {Hemley}},\ }\href@noop {}
	{\bibfield  {journal} {\bibinfo  {journal} {PNAS}\ }\textbf {\bibinfo
			{volume} {114}},\ \bibinfo {pages} {6990} (\bibinfo {year}
		{2017})}\BibitemShut {NoStop}%
	\bibitem [{\citenamefont {Semenok}\ \emph
		{et~al.}(2020{\natexlab{b}})\citenamefont {Semenok}, \citenamefont {Kvashin},
		\citenamefont {Ivanova}, \citenamefont {Svitlyk}, \citenamefont {Fominski},
		\citenamefont {Sadakov}, \citenamefont {Sobolevskiy}, \citenamefont
		{Pudalov}, \citenamefont {Troyan},\ and\ \citenamefont
		{Oganov}}]{Oganov_MatToday_2020_ThH}%
	\BibitemOpen
	\bibfield  {author} {\bibinfo {author} {\bibfnamefont {D.~V.}\ \bibnamefont
			{Semenok}}, \bibinfo {author} {\bibfnamefont {A.~G.}\ \bibnamefont
			{Kvashin}}, \bibinfo {author} {\bibfnamefont {A.~G.}\ \bibnamefont
			{Ivanova}}, \bibinfo {author} {\bibfnamefont {V.}~\bibnamefont {Svitlyk}},
		\bibinfo {author} {\bibfnamefont {V.~Y.}\ \bibnamefont {Fominski}}, \bibinfo
		{author} {\bibfnamefont {A.~V.}\ \bibnamefont {Sadakov}}, \bibinfo {author}
		{\bibfnamefont {O.~A.}\ \bibnamefont {Sobolevskiy}}, \bibinfo {author}
		{\bibfnamefont {V.~M.}\ \bibnamefont {Pudalov}}, \bibinfo {author}
		{\bibfnamefont {I.~A.}\ \bibnamefont {Troyan}}, \ and\ \bibinfo {author}
		{\bibfnamefont {A.~R.}\ \bibnamefont {Oganov}},\ }\href@noop {} {\bibfield
		{journal} {\bibinfo  {journal} {Materials Today}\ }\textbf {\bibinfo {volume}
			{33}},\ \bibinfo {pages} {36} (\bibinfo {year}
		{2020}{\natexlab{b}})}\BibitemShut {NoStop}%
	\bibitem [{\citenamefont {Troyan}\ \emph {et~al.}(2019)\citenamefont {Troyan},
		\citenamefont {Semenok}, \citenamefont {Kvashin}, \citenamefont {Sadakov},
		\citenamefont {Sobolevskiy}, \citenamefont {Pudalov}, \citenamefont
		{Ivanova}, \citenamefont {Prakapenka}, \citenamefont {Greenberg},
		\citenamefont {Gavriliuk}, \citenamefont {Struzhkin}, \citenamefont
		{Bergara}, \citenamefont {Errea}, \citenamefont {Bianco}, \citenamefont
		{Calandra}, \citenamefont {Mauri}, \citenamefont {Monacelli}, \citenamefont
		{Akashi},\ and\ \citenamefont {Oganov}}]{Oganov_arXiv_2019_YH6}%
	\BibitemOpen
	\bibfield  {author} {\bibinfo {author} {\bibfnamefont {I.~A.}\ \bibnamefont
			{Troyan}}, \bibinfo {author} {\bibfnamefont {D.~V.}\ \bibnamefont {Semenok}},
		\bibinfo {author} {\bibfnamefont {A.~G.}\ \bibnamefont {Kvashin}}, \bibinfo
		{author} {\bibfnamefont {A.~V.}\ \bibnamefont {Sadakov}}, \bibinfo {author}
		{\bibfnamefont {O.~A.}\ \bibnamefont {Sobolevskiy}}, \bibinfo {author}
		{\bibfnamefont {V.~M.}\ \bibnamefont {Pudalov}}, \bibinfo {author}
		{\bibfnamefont {A.~G.}\ \bibnamefont {Ivanova}}, \bibinfo {author}
		{\bibfnamefont {V.~B.}\ \bibnamefont {Prakapenka}}, \bibinfo {author}
		{\bibfnamefont {E.}~\bibnamefont {Greenberg}}, \bibinfo {author}
		{\bibfnamefont {A.~G.}\ \bibnamefont {Gavriliuk}}, \bibinfo {author}
		{\bibfnamefont {V.~V.}\ \bibnamefont {Struzhkin}}, \bibinfo {author}
		{\bibfnamefont {A.}~\bibnamefont {Bergara}}, \bibinfo {author} {\bibfnamefont
			{I.}~\bibnamefont {Errea}}, \bibinfo {author} {\bibfnamefont
			{R.}~\bibnamefont {Bianco}}, \bibinfo {author} {\bibfnamefont
			{M.}~\bibnamefont {Calandra}}, \bibinfo {author} {\bibfnamefont
			{F.}~\bibnamefont {Mauri}}, \bibinfo {author} {\bibfnamefont
			{L.}~\bibnamefont {Monacelli}}, \bibinfo {author} {\bibfnamefont
			{R.}~\bibnamefont {Akashi}}, \ and\ \bibinfo {author} {\bibfnamefont {A.~R.}\
			\bibnamefont {Oganov}},\ }\href@noop {} {\bibfield  {journal} {\bibinfo
			{journal} {arXiv:1908.01534}\ } (\bibinfo {year} {2019})}\BibitemShut
	{NoStop}%
	\bibitem [{\citenamefont {Kong}\ \emph {et~al.}(2019)\citenamefont {Kong},
		\citenamefont {Minkov}, \citenamefont {Kuzonikov}, \citenamefont {Besedin},
		\citenamefont {Drodzov}, \citenamefont {Mozaffari}, \citenamefont {Balicas},
		\citenamefont {Balakirev}, \citenamefont {Prakapenka}, \citenamefont
		{Greenberg}, \citenamefont {Knyazev},\ and\ \citenamefont
		{Eremets}}]{Eremets_arXiv_2019_YH6}%
	\BibitemOpen
	\bibfield  {author} {\bibinfo {author} {\bibfnamefont {P.~P.}\ \bibnamefont
			{Kong}}, \bibinfo {author} {\bibfnamefont {V.~S.}\ \bibnamefont {Minkov}},
		\bibinfo {author} {\bibfnamefont {M.~A.}\ \bibnamefont {Kuzonikov}}, \bibinfo
		{author} {\bibfnamefont {S.~P.}\ \bibnamefont {Besedin}}, \bibinfo {author}
		{\bibfnamefont {A.~P.}\ \bibnamefont {Drodzov}}, \bibinfo {author}
		{\bibfnamefont {S.}~\bibnamefont {Mozaffari}}, \bibinfo {author}
		{\bibfnamefont {L.}~\bibnamefont {Balicas}}, \bibinfo {author} {\bibfnamefont
			{F.~F.}\ \bibnamefont {Balakirev}}, \bibinfo {author} {\bibfnamefont {V.~B.}\
			\bibnamefont {Prakapenka}}, \bibinfo {author} {\bibfnamefont
			{E.}~\bibnamefont {Greenberg}}, \bibinfo {author} {\bibfnamefont {D.~A.}\
			\bibnamefont {Knyazev}}, \ and\ \bibinfo {author} {\bibfnamefont {M.~I.}\
			\bibnamefont {Eremets}},\ }\href@noop {} {\bibfield  {journal} {\bibinfo
			{journal} {arXiv:1909.10482}\ } (\bibinfo {year} {2019})}\BibitemShut
	{NoStop}%
	\bibitem [{\citenamefont {Drodzov}\ \emph
		{et~al.}(2015{\natexlab{b}})\citenamefont {Drodzov}, \citenamefont
		{Eremets},\ and\ \citenamefont {Troyan}}]{Eremets_arXiv_PH3_2015}%
	\BibitemOpen
	\bibfield  {author} {\bibinfo {author} {\bibfnamefont {A.~P.}\ \bibnamefont
			{Drodzov}}, \bibinfo {author} {\bibfnamefont {M.~I.}\ \bibnamefont
			{Eremets}}, \ and\ \bibinfo {author} {\bibfnamefont {I.~A.}\ \bibnamefont
			{Troyan}},\ }\href@noop {} {\bibfield  {journal} {\bibinfo  {journal}
			{arXiv:1508.06224}\ } (\bibinfo {year} {2015}{\natexlab{b}})}\BibitemShut
	{NoStop}%
	\bibitem [{\citenamefont {Snider}\ \emph {et~al.}(2020)\citenamefont {Snider},
		\citenamefont {Dasenbrock-Gammon}, \citenamefont {McBride}, \citenamefont
		{Debessai}, \citenamefont {Vindana}, \citenamefont {Vencatasamy},
		\citenamefont {Lawler}, \citenamefont {Salamat},\ and\ \citenamefont
		{Dias}}]{Dias_Nature_2020_CSH}%
	\BibitemOpen
	\bibfield  {author} {\bibinfo {author} {\bibfnamefont {E.}~\bibnamefont
			{Snider}}, \bibinfo {author} {\bibfnamefont {N.}~\bibnamefont
			{Dasenbrock-Gammon}}, \bibinfo {author} {\bibfnamefont {R.}~\bibnamefont
			{McBride}}, \bibinfo {author} {\bibfnamefont {M.}~\bibnamefont {Debessai}},
		\bibinfo {author} {\bibfnamefont {H.}~\bibnamefont {Vindana}}, \bibinfo
		{author} {\bibfnamefont {K.}~\bibnamefont {Vencatasamy}}, \bibinfo {author}
		{\bibfnamefont {K.~V.}\ \bibnamefont {Lawler}}, \bibinfo {author}
		{\bibfnamefont {A.}~\bibnamefont {Salamat}}, \ and\ \bibinfo {author}
		{\bibfnamefont {R.~P.}\ \bibnamefont {Dias}},\ }\href@noop {} {\bibfield
		{journal} {\bibinfo  {journal} {Nature}\ }\textbf {\bibinfo {volume} {586}},\
		\bibinfo {pages} {373} (\bibinfo {year} {2020})}\BibitemShut {NoStop}%
	\bibitem [{\citenamefont {Sun}\ \emph {et~al.}(2019{\natexlab{a}})\citenamefont
		{Sun}, \citenamefont {Lv}, \citenamefont {Xie}, \citenamefont {Liu},\ and\
		\citenamefont {Ma}}]{Ma_PRL_2019_Li2MgH16}%
	\BibitemOpen
	\bibfield  {author} {\bibinfo {author} {\bibfnamefont {Y.}~\bibnamefont
			{Sun}}, \bibinfo {author} {\bibfnamefont {J.}~\bibnamefont {Lv}}, \bibinfo
		{author} {\bibfnamefont {Y.}~\bibnamefont {Xie}}, \bibinfo {author}
		{\bibfnamefont {H.}~\bibnamefont {Liu}}, \ and\ \bibinfo {author}
		{\bibfnamefont {Y.}~\bibnamefont {Ma}},\ }\href@noop {} {\bibfield  {journal}
		{\bibinfo  {journal} {Phys. Rev. Lett.}\ }\textbf {\bibinfo {volume} {123}},\
		\bibinfo {pages} {097001} (\bibinfo {year} {2019}{\natexlab{a}})}\BibitemShut
	{NoStop}%
	\bibitem [{\citenamefont {Grockowiak}\ \emph {et~al.}(2020)\citenamefont
		{Grockowiak}, \citenamefont {Ahart}, \citenamefont {Helm}, \citenamefont
		{Coniglio}, \citenamefont {Kumar}, \citenamefont {Somayazulu}, \citenamefont
		{Meng}, \citenamefont {Oliff}, \citenamefont {Williams}, \citenamefont
		{Ashcroft}, \citenamefont {Hemley},\ and\ \citenamefont
		{Tozer}}]{superconductivity_550K}%
	\BibitemOpen
	\bibfield  {author} {\bibinfo {author} {\bibfnamefont {A.~D.}\ \bibnamefont
			{Grockowiak}}, \bibinfo {author} {\bibfnamefont {M.}~\bibnamefont {Ahart}},
		\bibinfo {author} {\bibfnamefont {T.}~\bibnamefont {Helm}}, \bibinfo {author}
		{\bibfnamefont {W.~A.}\ \bibnamefont {Coniglio}}, \bibinfo {author}
		{\bibfnamefont {R.}~\bibnamefont {Kumar}}, \bibinfo {author} {\bibfnamefont
			{M.}~\bibnamefont {Somayazulu}}, \bibinfo {author} {\bibfnamefont
			{Y.}~\bibnamefont {Meng}}, \bibinfo {author} {\bibfnamefont {M.}~\bibnamefont
			{Oliff}}, \bibinfo {author} {\bibfnamefont {V.}~\bibnamefont {Williams}},
		\bibinfo {author} {\bibfnamefont {N.~W.}\ \bibnamefont {Ashcroft}}, \bibinfo
		{author} {\bibfnamefont {R.~J.}\ \bibnamefont {Hemley}}, \ and\ \bibinfo
		{author} {\bibfnamefont {S.~W.}\ \bibnamefont {Tozer}},\ }\href@noop {}
	{\bibfield  {journal} {\bibinfo  {journal} {arXiv:2006.03004}\ } (\bibinfo
		{year} {2020})}\BibitemShut {NoStop}%
	\bibitem [{\citenamefont {Boeri}\ and\ \citenamefont
		{Bachelet}(2019)}]{Boeri_JPCM_2019_viewpoint}%
	\BibitemOpen
	\bibfield  {author} {\bibinfo {author} {\bibfnamefont {L.}~\bibnamefont
			{Boeri}}\ and\ \bibinfo {author} {\bibfnamefont {G.~B.}\ \bibnamefont
			{Bachelet}},\ }\href@noop {} {\bibfield  {journal} {\bibinfo  {journal} {J.
				Phys.: Condens. Matter}\ }\textbf {\bibinfo {volume} {31}},\ \bibinfo {pages}
		{234002} (\bibinfo {year} {2019})}\BibitemShut {NoStop}%
	\bibitem [{\citenamefont {Errea}\ \emph {et~al.}(2016)\citenamefont {Errea},
		\citenamefont {Calandra}, \citenamefont {Pickard}, \citenamefont {Nelson},
		\citenamefont {Needs}, \citenamefont {Li}, \citenamefont {Liu}, \citenamefont
		{Zhang}, \citenamefont {Ma},\ and\ \citenamefont
		{Mauri}}]{Mauri_Nature_2016_SH3}%
	\BibitemOpen
	\bibfield  {author} {\bibinfo {author} {\bibfnamefont {I.}~\bibnamefont
			{Errea}}, \bibinfo {author} {\bibfnamefont {M.}~\bibnamefont {Calandra}},
		\bibinfo {author} {\bibfnamefont {C.~J.}\ \bibnamefont {Pickard}}, \bibinfo
		{author} {\bibfnamefont {J.~R.}\ \bibnamefont {Nelson}}, \bibinfo {author}
		{\bibfnamefont {R.~J.}\ \bibnamefont {Needs}}, \bibinfo {author}
		{\bibfnamefont {Y.}~\bibnamefont {Li}}, \bibinfo {author} {\bibfnamefont
			{H.}~\bibnamefont {Liu}}, \bibinfo {author} {\bibfnamefont {Y.}~\bibnamefont
			{Zhang}}, \bibinfo {author} {\bibfnamefont {Y.}~\bibnamefont {Ma}}, \ and\
		\bibinfo {author} {\bibfnamefont {F.}~\bibnamefont {Mauri}},\ }\href@noop {}
	{\bibfield  {journal} {\bibinfo  {journal} {Nature}\ }\textbf {\bibinfo
			{volume} {532}},\ \bibinfo {pages} {81} (\bibinfo {year} {2016})}\BibitemShut
	{NoStop}%
	\bibitem [{\citenamefont {Errea}\ \emph {et~al.}(2020)\citenamefont {Errea},
		\citenamefont {Belli}, \citenamefont {Monacelli}, \citenamefont {Sanna},
		\citenamefont {Koretsune}, \citenamefont {Tadano}, \citenamefont {Bianco},
		\citenamefont {Calandra}, \citenamefont {Arita}, \citenamefont {Mauri},\ and\
		\citenamefont {Flores-Livas}}]{Mauri_Nature_2020_LaH}%
	\BibitemOpen
	\bibfield  {author} {\bibinfo {author} {\bibfnamefont {I.}~\bibnamefont
			{Errea}}, \bibinfo {author} {\bibfnamefont {F.}~\bibnamefont {Belli}},
		\bibinfo {author} {\bibfnamefont {L.}~\bibnamefont {Monacelli}}, \bibinfo
		{author} {\bibfnamefont {A.}~\bibnamefont {Sanna}}, \bibinfo {author}
		{\bibfnamefont {T.}~\bibnamefont {Koretsune}}, \bibinfo {author}
		{\bibfnamefont {T.}~\bibnamefont {Tadano}}, \bibinfo {author} {\bibfnamefont
			{R.}~\bibnamefont {Bianco}}, \bibinfo {author} {\bibfnamefont
			{M.}~\bibnamefont {Calandra}}, \bibinfo {author} {\bibfnamefont
			{R.}~\bibnamefont {Arita}}, \bibinfo {author} {\bibfnamefont
			{F.}~\bibnamefont {Mauri}}, \ and\ \bibinfo {author} {\bibfnamefont {J.~A.}\
			\bibnamefont {Flores-Livas}},\ }\href@noop {} {\bibfield  {journal} {\bibinfo
			{journal} {Nature}\ }\textbf {\bibinfo {volume} {578}},\ \bibinfo {pages}
		{66} (\bibinfo {year} {2020})}\BibitemShut {NoStop}%
	\bibitem [{\citenamefont {Zurek}\ \emph {et~al.}(2009)\citenamefont {Zurek},
		\citenamefont {Hoffmann}, \citenamefont {Ashcroft}, \citenamefont {Oganov},\
		and\ \citenamefont {Lyakhov}}]{Zurek_PNAS_2009_LiH}%
	\BibitemOpen
	\bibfield  {author} {\bibinfo {author} {\bibfnamefont {E.}~\bibnamefont
			{Zurek}}, \bibinfo {author} {\bibfnamefont {R.}~\bibnamefont {Hoffmann}},
		\bibinfo {author} {\bibfnamefont {N.~W.}\ \bibnamefont {Ashcroft}}, \bibinfo
		{author} {\bibfnamefont {A.~R.}\ \bibnamefont {Oganov}}, \ and\ \bibinfo
		{author} {\bibfnamefont {A.~O.}\ \bibnamefont {Lyakhov}},\ }\href@noop {}
	{\bibfield  {journal} {\bibinfo  {journal} {PNAS}\ }\textbf {\bibinfo
			{volume} {106}},\ \bibinfo {pages} {17640} (\bibinfo {year}
		{2009})}\BibitemShut {NoStop}%
	\bibitem [{\citenamefont {Peng}\ \emph {et~al.}(2017)\citenamefont {Peng},
		\citenamefont {Sun}, \citenamefont {Pickard}, \citenamefont {Needs},
		\citenamefont {Wu},\ and\ \citenamefont {Ma}}]{Ma_PRL_2017_ReH}%
	\BibitemOpen
	\bibfield  {author} {\bibinfo {author} {\bibfnamefont {F.}~\bibnamefont
			{Peng}}, \bibinfo {author} {\bibfnamefont {Y.}~\bibnamefont {Sun}}, \bibinfo
		{author} {\bibfnamefont {C.~J.}\ \bibnamefont {Pickard}}, \bibinfo {author}
		{\bibfnamefont {R.~J.}\ \bibnamefont {Needs}}, \bibinfo {author}
		{\bibfnamefont {Q.}~\bibnamefont {Wu}}, \ and\ \bibinfo {author}
		{\bibfnamefont {Y.}~\bibnamefont {Ma}},\ }\href@noop {} {\bibfield  {journal}
		{\bibinfo  {journal} {Phys. Rev. Lett.}\ }\textbf {\bibinfo {volume} {119}},\
		\bibinfo {pages} {107001} (\bibinfo {year} {2017})}\BibitemShut {NoStop}%
	\bibitem [{\citenamefont {Glass}\ \emph {et~al.}(2006)\citenamefont {Glass},
		\citenamefont {Oganov},\ and\ \citenamefont {Hansen}}]{USPEX_1}%
	\BibitemOpen
	\bibfield  {author} {\bibinfo {author} {\bibfnamefont {C.~W.}\ \bibnamefont
			{Glass}}, \bibinfo {author} {\bibfnamefont {A.~R.}\ \bibnamefont {Oganov}}, \
		and\ \bibinfo {author} {\bibfnamefont {N.}~\bibnamefont {Hansen}},\
	}\href@noop {} {\bibfield  {journal} {\bibinfo  {journal} {Computer Physics
				Communication}\ }\textbf {\bibinfo {volume} {175}},\ \bibinfo {pages} {713}
		(\bibinfo {year} {2006})}\BibitemShut {NoStop}%
	\bibitem [{\citenamefont {Lyakhov}\ \emph {et~al.}(2013)\citenamefont
		{Lyakhov}, \citenamefont {Oganov}, \citenamefont {Stokes},\ and\
		\citenamefont {Zhu}}]{USPEX_2}%
	\BibitemOpen
	\bibfield  {author} {\bibinfo {author} {\bibfnamefont {A.~O.}\ \bibnamefont
			{Lyakhov}}, \bibinfo {author} {\bibfnamefont {A.~R.}\ \bibnamefont {Oganov}},
		\bibinfo {author} {\bibfnamefont {H.~T.}\ \bibnamefont {Stokes}}, \ and\
		\bibinfo {author} {\bibfnamefont {Q.}~\bibnamefont {Zhu}},\ }\href@noop {}
	{\bibfield  {journal} {\bibinfo  {journal} {Computer Physics Communication}\
		}\textbf {\bibinfo {volume} {184}},\ \bibinfo {pages} {1172} (\bibinfo {year}
		{2013})}\BibitemShut {NoStop}%
	\bibitem [{\citenamefont {Sun}\ \emph {et~al.}(2019{\natexlab{b}})\citenamefont
		{Sun}, \citenamefont {Bartel}, \citenamefont {Arca}, \citenamefont {Bauers},
		\citenamefont {Matthews}, \citenamefont {Orvananos}, \citenamefont {Chen},
		\citenamefont {Toney}, \citenamefont {Schelhas}, \citenamefont {Tumas},
		\citenamefont {Tate}, \citenamefont {Zakutayev}, \citenamefont {Lany},
		\citenamefont {Holder},\ and\ \citenamefont {Ceder}}]{Ceder_NatMat_2019}%
	\BibitemOpen
	\bibfield  {author} {\bibinfo {author} {\bibfnamefont {W.}~\bibnamefont
			{Sun}}, \bibinfo {author} {\bibfnamefont {C.~J.}\ \bibnamefont {Bartel}},
		\bibinfo {author} {\bibfnamefont {E.}~\bibnamefont {Arca}}, \bibinfo {author}
		{\bibfnamefont {S.~R.}\ \bibnamefont {Bauers}}, \bibinfo {author}
		{\bibfnamefont {B.}~\bibnamefont {Matthews}}, \bibinfo {author}
		{\bibfnamefont {B.}~\bibnamefont {Orvananos}}, \bibinfo {author}
		{\bibfnamefont {B.-R.}\ \bibnamefont {Chen}}, \bibinfo {author}
		{\bibfnamefont {M.~F.}\ \bibnamefont {Toney}}, \bibinfo {author}
		{\bibfnamefont {L.~T.}\ \bibnamefont {Schelhas}}, \bibinfo {author}
		{\bibfnamefont {W.}~\bibnamefont {Tumas}}, \bibinfo {author} {\bibfnamefont
			{J.}~\bibnamefont {Tate}}, \bibinfo {author} {\bibfnamefont {A.}~\bibnamefont
			{Zakutayev}}, \bibinfo {author} {\bibfnamefont {S.}~\bibnamefont {Lany}},
		\bibinfo {author} {\bibfnamefont {A.~M.}\ \bibnamefont {Holder}}, \ and\
		\bibinfo {author} {\bibfnamefont {G.}~\bibnamefont {Ceder}},\ }\href@noop {}
	{\bibfield  {journal} {\bibinfo  {journal} {Nature Materials}\ }\textbf
		{\bibinfo {volume} {18}} (\bibinfo {year} {2019}{\natexlab{b}})}\BibitemShut
	{NoStop}%
	\bibitem [{\citenamefont {Teredesai}\ \emph {et~al.}(2004)\citenamefont
		{Teredesai}, \citenamefont {Muthu}, \citenamefont {Chandrabhas},
		\citenamefont {Meenakshi}, \citenamefont {Vijayakumar}, \citenamefont
		{Modak}, \citenamefont {Rao}, \citenamefont {Godwal}, \citenamefont {Sikka},\
		and\ \citenamefont {Sood}}]{Sood_SSC_2004_LaB6}%
	\BibitemOpen
	\bibfield  {author} {\bibinfo {author} {\bibfnamefont {P.}~\bibnamefont
			{Teredesai}}, \bibinfo {author} {\bibfnamefont {D.~V.~S.}\ \bibnamefont
			{Muthu}}, \bibinfo {author} {\bibfnamefont {N.}~\bibnamefont {Chandrabhas}},
		\bibinfo {author} {\bibfnamefont {S.}~\bibnamefont {Meenakshi}}, \bibinfo
		{author} {\bibfnamefont {V.}~\bibnamefont {Vijayakumar}}, \bibinfo {author}
		{\bibfnamefont {P.}~\bibnamefont {Modak}}, \bibinfo {author} {\bibfnamefont
			{R.~S.}\ \bibnamefont {Rao}}, \bibinfo {author} {\bibfnamefont {B.~K.}\
			\bibnamefont {Godwal}}, \bibinfo {author} {\bibfnamefont {S.~K.}\
			\bibnamefont {Sikka}}, \ and\ \bibinfo {author} {\bibfnamefont {A.~K.}\
			\bibnamefont {Sood}},\ }\href@noop {} {\bibfield  {journal} {\bibinfo
			{journal} {Solid State Comm.}\ }\textbf {\bibinfo {volume} {129}} (\bibinfo
		{year} {2004})}\BibitemShut {NoStop}%
	\bibitem [{\citenamefont {Cava}\ \emph {et~al.}(1994)\citenamefont {Cava},
		\citenamefont {Zandbergen}, \citenamefont {Batlogg}, \citenamefont {Eisaki},
		\citenamefont {Tagaki}, \citenamefont {Krajewski}, \citenamefont {Peck},
		\citenamefont {Gyorgy},\ and\ \citenamefont
		{Uchida}}]{Uchida_Nature_1994_LaNiBN}%
	\BibitemOpen
	\bibfield  {author} {\bibinfo {author} {\bibfnamefont {R.~J.}\ \bibnamefont
			{Cava}}, \bibinfo {author} {\bibfnamefont {H.~W.}\ \bibnamefont
			{Zandbergen}}, \bibinfo {author} {\bibfnamefont {B.}~\bibnamefont {Batlogg}},
		\bibinfo {author} {\bibfnamefont {H.}~\bibnamefont {Eisaki}}, \bibinfo
		{author} {\bibfnamefont {H.}~\bibnamefont {Tagaki}}, \bibinfo {author}
		{\bibfnamefont {J.~J.}\ \bibnamefont {Krajewski}}, \bibinfo {author}
		{\bibfnamefont {W.~F.}\ \bibnamefont {Peck}}, \bibinfo {author}
		{\bibfnamefont {E.~M.}\ \bibnamefont {Gyorgy}}, \ and\ \bibinfo {author}
		{\bibfnamefont {S.}~\bibnamefont {Uchida}},\ }\href@noop {} {\bibfield
		{journal} {\bibinfo  {journal} {Nature}\ }\textbf {\bibinfo {volume} {273}},\
		\bibinfo {pages} {245} (\bibinfo {year} {1994})}\BibitemShut {NoStop}%
	\bibitem [{\citenamefont {Cataldo}\ \emph {et~al.}(2020)\citenamefont
		{Cataldo}, \citenamefont {von~der Linden},\ and\ \citenamefont
		{Boeri}}]{DiCataldo_PRB_2020_CaBH}%
	\BibitemOpen
	\bibfield  {author} {\bibinfo {author} {\bibfnamefont {S.~D.}\ \bibnamefont
			{Cataldo}}, \bibinfo {author} {\bibfnamefont {W.}~\bibnamefont {von~der
				Linden}}, \ and\ \bibinfo {author} {\bibfnamefont {L.}~\bibnamefont
			{Boeri}},\ }\href@noop {} {\bibfield  {journal} {\bibinfo  {journal} {Phys.
				Rev. B}\ }\textbf {\bibinfo {volume} {102}},\ \bibinfo {pages} {014516}
		(\bibinfo {year} {2020})}\BibitemShut {NoStop}%
	\bibitem [{\citenamefont {Kokail}\ \emph {et~al.}(2017)\citenamefont {Kokail},
		\citenamefont {von~der Linden},\ and\ \citenamefont
		{Boeri}}]{Kokail_PRM_2017_LiBH}%
	\BibitemOpen
	\bibfield  {author} {\bibinfo {author} {\bibfnamefont {C.}~\bibnamefont
			{Kokail}}, \bibinfo {author} {\bibfnamefont {W.}~\bibnamefont {von~der
				Linden}}, \ and\ \bibinfo {author} {\bibfnamefont {L.}~\bibnamefont
			{Boeri}},\ }\href@noop {} {\bibfield  {journal} {\bibinfo  {journal} {Phys.
				Rev. M}\ }\textbf {\bibinfo {volume} {1}},\ \bibinfo {pages} {074803}
		(\bibinfo {year} {2017})}\BibitemShut {NoStop}%
	\bibitem [{Note1()}]{Note1}%
	\BibitemOpen
	\bibinfo {note} {The Supplemental Material is available at..}\BibitemShut
	{Stop}%
	\bibitem [{\citenamefont {Cataldo}\ \emph {et~al.}(2021)\citenamefont
		{Cataldo}, \citenamefont {Heil}, \citenamefont {von~der Linden},\ and\
		\citenamefont {Boeri}}]{DiCataldo_arXiv_LaBH8_2021}%
	\BibitemOpen
	\bibfield  {author} {\bibinfo {author} {\bibfnamefont {S.~D.}\ \bibnamefont
			{Cataldo}}, \bibinfo {author} {\bibfnamefont {C.}~\bibnamefont {Heil}},
		\bibinfo {author} {\bibfnamefont {W.}~\bibnamefont {von~der Linden}}, \ and\
		\bibinfo {author} {\bibfnamefont {L.}~\bibnamefont {Boeri}},\ }\href@noop {}
	{\bibfield  {journal} {\bibinfo  {journal} {arXiv:2102.11227}\ } (\bibinfo
		{year} {2021})}\BibitemShut {NoStop}%
	\bibitem [{\citenamefont {Belli}\ \emph {et~al.}(2021)\citenamefont {Belli},
		\citenamefont {Contreras-Garcia},\ and\ \citenamefont
		{Errea}}]{Errea_arXiv_2021_hydridebased}%
	\BibitemOpen
	\bibfield  {author} {\bibinfo {author} {\bibfnamefont {F.}~\bibnamefont
			{Belli}}, \bibinfo {author} {\bibfnamefont {J.}~\bibnamefont
			{Contreras-Garcia}}, \ and\ \bibinfo {author} {\bibfnamefont
			{I.}~\bibnamefont {Errea}},\ }\href@noop {} {\bibfield  {journal} {\bibinfo
			{journal} {arXiv:2103.07320}\ } (\bibinfo {year} {2021})}\BibitemShut
	{NoStop}%
	\bibitem [{\citenamefont {Borinaga}\ \emph {et~al.}(2016)\citenamefont
		{Borinaga}, \citenamefont {Errea}, \citenamefont {Calandra}, \citenamefont
		{Mauri},\ and\ \citenamefont {Bergara}}]{Borinaga_PRB_2016_atomicH}%
	\BibitemOpen
	\bibfield  {author} {\bibinfo {author} {\bibfnamefont {M.}~\bibnamefont
			{Borinaga}}, \bibinfo {author} {\bibfnamefont {I.}~\bibnamefont {Errea}},
		\bibinfo {author} {\bibfnamefont {M.}~\bibnamefont {Calandra}}, \bibinfo
		{author} {\bibfnamefont {F.}~\bibnamefont {Mauri}}, \ and\ \bibinfo {author}
		{\bibfnamefont {A.}~\bibnamefont {Bergara}},\ }\href@noop {} {\bibfield
		{journal} {\bibinfo  {journal} {Phys. Rev. B}\ }\textbf {\bibinfo {volume}
			{93}},\ \bibinfo {pages} {174308} (\bibinfo {year} {2016})}\BibitemShut
	{NoStop}%
	\bibitem [{\citenamefont {Azadi}\ \emph {et~al.}(2014)\citenamefont {Azadi},
		\citenamefont {Monserrat}, \citenamefont {Foulkes},\ and\ \citenamefont
		{Needs}}]{Needs_PRL_2014_atomicH}%
	\BibitemOpen
	\bibfield  {author} {\bibinfo {author} {\bibfnamefont {S.}~\bibnamefont
			{Azadi}}, \bibinfo {author} {\bibfnamefont {B.}~\bibnamefont {Monserrat}},
		\bibinfo {author} {\bibfnamefont {W.~M.~C.}\ \bibnamefont {Foulkes}}, \ and\
		\bibinfo {author} {\bibfnamefont {R.~J.}\ \bibnamefont {Needs}},\ }\href@noop
	{} {\bibfield  {journal} {\bibinfo  {journal} {Phys. Rev. Lett.}\ }\textbf
		{\bibinfo {volume} {112}},\ \bibinfo {pages} {165501} (\bibinfo {year}
		{2014})}\BibitemShut {NoStop}%
	\bibitem [{\citenamefont {Allen}\ and\ \citenamefont
		{Dynes}(1975)}]{Allen_PRB_1975_McMillan}%
	\BibitemOpen
	\bibfield  {author} {\bibinfo {author} {\bibfnamefont {P.~B.}\ \bibnamefont
			{Allen}}\ and\ \bibinfo {author} {\bibfnamefont {R.~C.}\ \bibnamefont
			{Dynes}},\ }\href@noop {} {\bibfield  {journal} {\bibinfo  {journal} {Phys.
				Rev. B}\ }\textbf {\bibinfo {volume} {12}},\ \bibinfo {pages} {905} (\bibinfo
		{year} {1975})}\BibitemShut {NoStop}%
	\bibitem [{\citenamefont {Carbotte}(1990)}]{Carbotte_RevModPhys_1990_sc}%
	\BibitemOpen
	\bibfield  {author} {\bibinfo {author} {\bibfnamefont {J.~P.}\ \bibnamefont
			{Carbotte}},\ }\href@noop {} {\bibfield  {journal} {\bibinfo  {journal} {Rev.
				Mod. Phys}\ }\textbf {\bibinfo {volume} {62}},\ \bibinfo {pages} {1027}
		(\bibinfo {year} {1990})}\BibitemShut {NoStop}%
	\bibitem [{\citenamefont {Heil}\ \emph {et~al.}(2019)\citenamefont {Heil},
		\citenamefont {Cataldo}, \citenamefont {Bachelet},\ and\ \citenamefont
		{Boeri}}]{Heil_PRB_2019}%
	\BibitemOpen
	\bibfield  {author} {\bibinfo {author} {\bibfnamefont {C.}~\bibnamefont
			{Heil}}, \bibinfo {author} {\bibfnamefont {S.~D.}\ \bibnamefont {Cataldo}},
		\bibinfo {author} {\bibfnamefont {G.~B.}\ \bibnamefont {Bachelet}}, \ and\
		\bibinfo {author} {\bibfnamefont {L.}~\bibnamefont {Boeri}},\ }\href@noop {}
	{\bibfield  {journal} {\bibinfo  {journal} {Phys. Rev. B}\ }\textbf {\bibinfo
			{volume} {99}},\ \bibinfo {pages} {220502(R)} (\bibinfo {year}
		{2019})}\BibitemShut {NoStop}%
\end{thebibliography}

%

\end{document}